\documentclass[fleqn,11pt]{wlscirep}
\usepackage[utf8]{inputenc}
\usepackage[T1]{fontenc}
\usepackage{hyperref}
\usepackage{authblk}
\usepackage{xcolor}
\usepackage{ragged2e}

\definecolor{myblue}{rgb}{0,0,255}
\definecolor{myred}{rgb}{255,0,0}
\title{A Foundation Model for Non-Destructive Defect Identification from Vibrational Spectra}

\author[1,2,3,$\dagger$,*]{Mouyang Cheng}
\author[1,4,$\dagger$]{Chu-Liang Fu}
\author[1,5,$\dagger$]{Bowen Yu}
\author[1,4]{Eunbi Rha}
\author[1,6]{Abhijatmedhi Chotrattanapituk}
\author[7]{Douglas L Abernathy}
\author[7]{Yongqiang Cheng}
\author[1,2,4,**]{Mingda Li}
\affil[1]{Quantum Measurement Group, MIT, Cambridge, MA 02139, USA}
\affil[2]{Center for Computational Science and Engineering, MIT, Cambridge, MA 02139, USA}
\affil[3]{Department of Materials Science and Engineering, MIT, Cambridge, MA 02139, USA}
\affil[4]{Department of Nuclear Science and Engineering, MIT, Cambridge, MA 02139, USA}
\affil[5]{Department of Physics, MIT, Cambridge, MA 02139, USA}
\affil[6]{Department of Electrical Engineering and Computer Science, MIT, Cambridge, MA 02139, USA}
\affil[7]{Neutron Scattering Division, Oak Ridge National Laboratory, Oak Ridge, TN, USA}

\affil[$\dagger$]{These authors contributed equally.}
\affil[*]{e-mail: vipandyc@mit.edu}
\affil[**]{e-mail: mingda@mit.edu}

\begin{abstract}
Defects are ubiquitous in solids and strongly influence materials’ mechanical and functional properties. However, non-destructive characterization and quantification of defects, especially when multiple types coexist, remain a long-standing challenge. Here we introduce DefectNet, a foundation machine learning model that predicts the chemical identity and concentration of substitutional point defects with multiple coexisting elements directly from vibrational spectra, specifically phonon density-of-states (PDoS). Trained on over 16,000 simulated spectra from 2,000 semiconductors, DefectNet employs a tailored attention mechanism to identify up to six distinct defect elements at concentrations ranging from 0.2\% to 25\%. The model generalizes well to unseen crystals across 56 elements and can be fine-tuned on experimental data. Validation using inelastic scattering measurements of SiGe alloys and MgB$_2$ superconductor demonstrates its accuracy and transferability. Our work establishes vibrational spectroscopy as a viable, non-destructive probe for point defect quantification in bulk materials, and highlights the promise of foundation models in data-driven defect engineering.

\end{abstract}
\begin{document}

\flushbottom
\maketitle
\thispagestyle{empty}

\section*{Introduction}
Defects play a pivotal role in materials science, profoundly influencing physical properties and ultimately determining device performance across a wide range of technological applications \cite{walsh2017instilling}. 
The atomic-scale defects are central to the functionality of materials, including solar cells \cite{ball2016defects}, thermoelectrics \cite{wu2024defect,fu2025ai}, electrode materials in batteries \cite{zhang2020defect}, catalysts \cite{xie2020defect}, and quantum materials \cite{wolfowicz2021quantum,awschalom2018quantum,sun2024unveiling}. 
Historically, the study of defects in semiconductors has received considerable attention since the 1950s, due to their critical impact on electronic and optical properties. For instance, controlled doping in silicon enables tunable carrier concentrations \cite{sze2021physics}, which are instrumental to transistor operation and microelectronics. Similarly, defect engineering in ultrawide bandgap semiconductors enhances their potential for next-generation power electronics \cite{tsao2018ultrawide}.
More recently, precise defect control, including doping, has emerged as a key enabler for quantum materials. In high-temperature superconductors, optimal superconductivity is achieved only within a range of hole-type doping concentrations \cite{orenstein2000advances,zhou2021high}.  
In flat-band systems, tuning the Fermi level via doping can unlock strongly correlated phases \cite{kang2020topological,checkelsky2024flat}.
Topological materials further highlight this sensitivity. In topological insulators and Weyl semimetals, quantum states such as the quantum anomalous Hall effect \cite{chang2013experimental,kang2023field} or large Berry curvature-induced responses \cite{li2020giant,han2020quantized} only manifest when the Fermi level is precisely tuned through defect or dopant engineering. 
The importance of point-defect engineering in functional materials, such as semiconductors and quantum materials, creates an urgent need for characterization methods that can accurately quantify the types and concentrations of defects.

However, despite the variety of defect characterization techniques, significant limitations remain in terms of sensitivity, selectivity, quantifiability, and destructiveness. For sensitivity, methods such as diffraction and diffuse scattering primarily respond to large-scale microstructures and inhomogeneities, but not to isolated point defects. Regarding selectivity, positron annihilation spectroscopy (PAS) is highly sensitive only to vacancy-type defects \cite{selim2021positron}. 
Techniques like deep-level transient spectroscopy (DLTS)\cite{lang1974deep}, nuclear magnetic resonance (NMR) \cite{rule2006nmr} and electron paramagnetic resonance (EPR) \cite{al2013electron} offer exceptional sensitivity to low-concentration defects, but are only selective to electrically active trap states (for DLTS), specific isotopes (for NMR), or without chemical specificity (EPR). 
As to quantification, optical probes such as photoluminescence (PL) \cite{bebb1972photoluminescence}, cathodoluminescence (CL) \cite{yacobi1990cathodoluminescence}, Raman and infrared (IR) spectroscopy \cite{schrader2008infrared}, along with X-ray probes like energy-dispersive X-ray spectroscopy (EDS), X-ray absorption spectroscopy (XAS) and X-ray photoelectron spectroscopy (XPS) \cite{de2005multiplet} are responsive to point defects, but often struggle to accurately quantify their concentrations, particularly in the presence of multiple coexisting defect species. 
Regarding destructiveness, imaging methods such as transmission electron microscopy (TEM) \cite{williams2009transmission} and atom probe tomography (APT) \cite{kelly2007atom} provide direct 2D and 3D visualization of defects, but require invasive sample preparation and are inherently destructive. 
Given these constraints, it would be highly desirable to develop characterization techniques that can accurately determine both the chemical identity and concentration of point defects with high sensitivity, broad chemical selectivity, and in a non-destructive manner.

In this work, we demonstrate that vibrational spectroscopy that measures PDoS, when integrated with machine learning, offers a promising route toward non-destructive characterization and quantification of point defects in bulk solids. 
We introduce DefectNet, a foundational machine learning model designed to infer both the chemical identity and concentration of point defects in solid-state materials from vibrational spectra. DefectNet takes as input the PDoS of pristine and defective crystals and outputs the defect chemical species and concentrations for up to six types of substitutional defects that coexist. To accommodate prior knowledge, DefectNet is flexibly designed to integrate initial guesses of candidate defect species, whether provided by the user or generated by an automated recommender. 
DefectNet addresses three key challenges. First, conventional atom-by-atom representations are inefficient across diverse materials with defects. We overcome this with the impurity-averaged configuration (IAC) representation, an encoding inspired by impurity average procedures in quantum field theory \cite{rammer1991quantum,bruus2004many,coleman2015introduction}. Second, at low concentrations, defective and pristine phonon spectra can appear nearly identical. To resolve this, we incorporate a spectral attention mechanism that enhances sensitivity to subtle perturbations. Third, the high cost of density functional theory (DFT)-based phonon simulations limits high-throughput data generation. We mitigate this by using machine-learning interatomic potentials (MLIPs) \cite{chen2022universal,deng2023chgnet,batatia2023foundation,yang2024mattersim} for efficient, high-fidelity vibrational spectra in broad chemical spaces. 
Trained on a dataset comprised of 16,000 simulated PDoS spectra derived from 2,000 distinct parent semiconductor materials, DefectNet accurately predicts defect types and concentrations as low as 0.2\%. We further demonstrate that DefectNet can be fine-tuned using experimental inelastic neutron scattering (INS) data to recover Ge concentrations in SiGe alloys, as well as Al doping in MgB$_2$ superconductors. 
DefectNet marks an initial step toward general-purpose, non-destructive defect characterization and quantification from vibrational spectra, enabling systematic characterization and guiding rational defect engineering through high-throughput simulation and machine learning.

\section*{Results}
\subsection*{Defect representation with impurity-averaged configuration}
We start by introducing the defect representations in our model. 
Inspired by the fact that defects are normally modeled as perturbations on the crystal structure, we split atomic embeddings into $\{f_{j\theta}\}$ for crystalline sites and $\{g_{j\alpha\theta}\}$ for defect sites.
Here, $j$ is the atom index, $\alpha$ represents the defect type, and $\theta$ indicates that all these embeddings are learnable features.
These features can be learned by performing an ensemble average on multiple atomic features. For example, for feature vector $\vec{f_j}=(1,Z_j,m_j,s_j,r_j,...)$, which may include physical quantities such as charge number $Z_j$, mass number $m_j$, spin $s_j$, and radius $r_j$, the atomic embedding can be expressed as 
\begin{equation}
    f_{j\theta} = \text{Embed}_\theta (1,Z_j,m_j,s_j,r_j,...)
\end{equation}
i.e., all input features of atom $j$ are optimized for their embedding mixture by an ensemble embedding layer $\text{Embed}_\theta$ \cite{Nguyen2024Optical}.
Similarly, $\vec{g}_{j\alpha}$ denotes the feature vector of a defect atom of type $\alpha$ at location $\mathbf{R}_{j\alpha}$, with the same embedding procedure as $\vec{f}_j$. 
Subsequently, to define the complete system with global geometry, we can define a generalized density operator with these atomic embeddings in Fourier space:
\begin{equation}
    \boldsymbol{\rho}(\mathbf{q}) = \boldsymbol{\rho}_{\mathrm{pristine}}(\mathbf{q}) + \boldsymbol{\rho}_{\mathrm{defect}}(\mathbf{q}) = \sum_j f_{j\theta}\, e^{i\mathbf{q} \cdot \mathbf{R}_{j}} +  \sum_{j\alpha} g_{j\alpha\theta}\, e^{i\mathbf{q} \cdot \mathbf{R}_{j\alpha}}.
    \label{rhoq}
\end{equation}

For the pristine part $\rho_{\text{pristine}}(\textbf{q})$, it captures the full structure information of the parent material. In our approach, we reduce it to the chemical formula and pristine PDoS, which shows surprisingly good performance even with significant dimensionality dimension compression. For the defect part $\rho_{\text{defect}}(\textbf{q})$, we introduce the Impurity-Averaged Configuration (IAC), a stochastic embedding of ensemble defect contributions, to provide a machine-readable defect representation. 
Rather than working with deterministic positions $\mathbf{R}_{j\alpha}$, we treat them as random variables drawn from a learnable distribution $P_\phi({\mathbf{R}_{j\alpha}})$, which encodes structural correlations such as short-range order (SRO) or spatial clustering. This yields a generalized form:
 \begin{equation}
\rho_{\text {IAC}}(\mathbf{q})=\overline{\left\langle\boldsymbol{\rho}_{\text {defect}}(\mathbf{q})\right\rangle}=\bigotimes_{j \alpha}\mathbb{E}_{\left\{\mathbf{R}_{j \alpha}\right\} \sim P_\phi}\left[g_{j\alpha\theta} \times e^{i \mathbf{q} \cdot \mathbf{R}_{j \alpha}}\right] .
     \label{eq:impurity_average}
 \end{equation}
Here $\bigotimes_{j \alpha}$ denotes a tensor product, and $\times$ represents elemental multiplication. The probability model $P_\phi$ can be explicitly parametrized via pairwise correlation functions, learned via deep neural networks, or sampled from known experimental distributions.
With the IAC defined by Eq.\,\ref{eq:impurity_average}, one can readily embed the general defect-related information through averaging all the detailed positional information, given the appropriate probability distribution $P_\phi$. 
Since the feature embedding $g_{j\alpha\theta}$ is introduced in general, the IAC can be applied to represent multiple configurations of correlated defects, including positional, charge, and spin correlations, which can even be further connected to general observables.

Assuming randomly and independently distributed defects, the IAC representation reduces to the lowest order, which is only dependent on defect density for uncorrelated defects:
\begin{align}
    \boldsymbol{\rho}_{\rm{IAC}}(\mathbf{q}) =  \bigoplus_\alpha n_{\alpha} \langle g_{j\alpha\theta} \rangle_j\, \delta_{\mathbf{q}0}= \bigoplus_\alpha n_{\alpha} g_{\alpha\theta}\, \delta_{\mathbf{q}0} 
\label{dopants}
\end{align}
where $n_{\alpha}$ is the density of type-$\alpha$ defects and $g_{\alpha\theta}$ is the average feature embedding for all atoms of that defect type.
Now if we focus on the first column of each feature embedding $g_{\alpha\theta}$, i.e. gives full weight of the first trivial atomic feature in Embed$_\theta$, the resulting IAC descriptor turns out to be a concatenated vector $\boldsymbol{\rho}_{\rm{IAC}}(\mathbf{q}) = (...,n_{\alpha},...)^T \delta_{\mathbf{q}0}$, which gives us exactly the dopant concentration of each defect $\alpha$ in a given material.
We note that the IAC representation above is inspired by the impurity average technique for defects in quantum field theory. Details regarding the physical motivation are discussed in Supplementary Information 1.

\subsection*{Model architecture of DefectNet}
\begin{figure}
  \centering
  \includegraphics[width=1.0\textwidth]{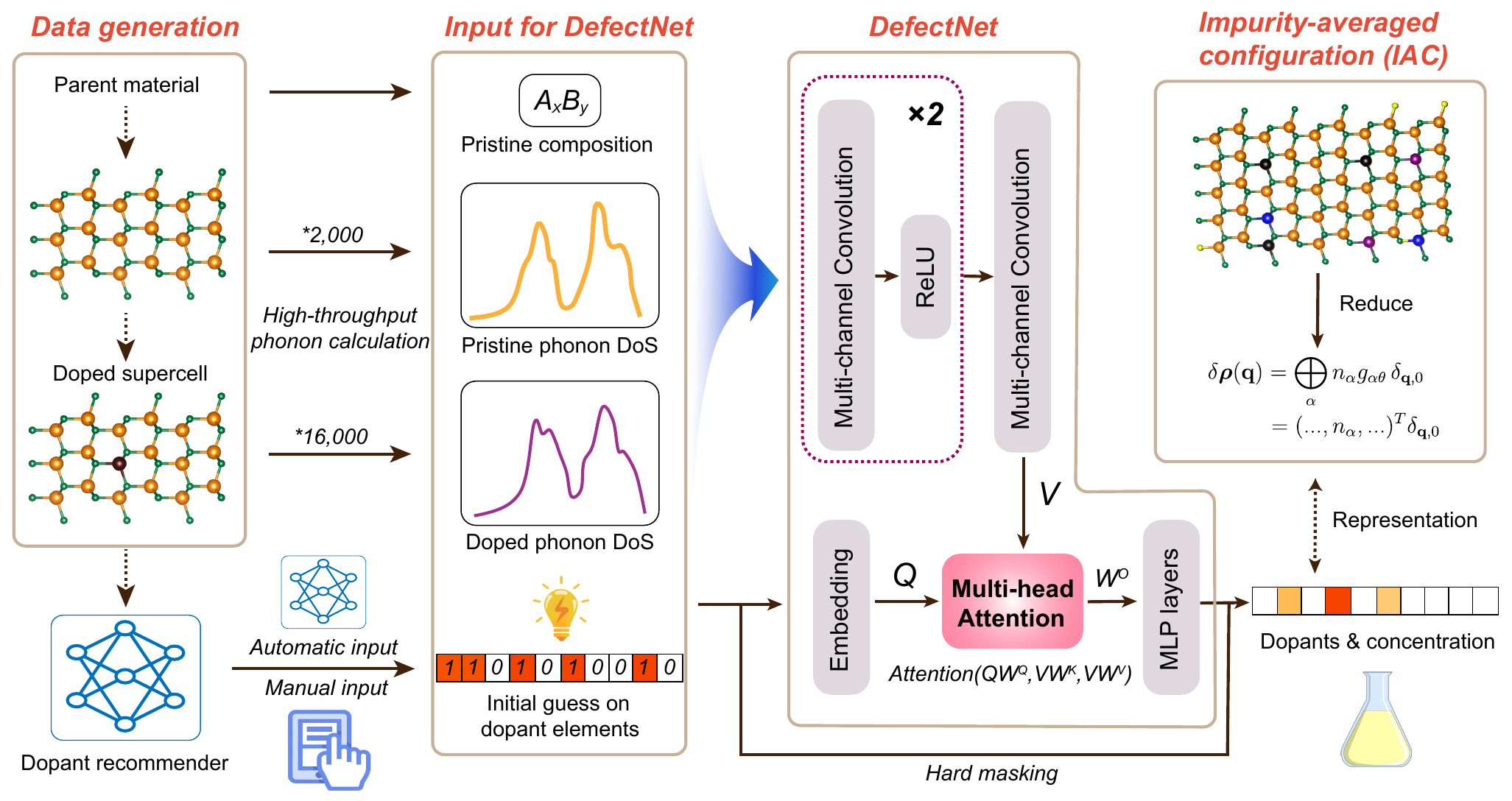}
  \caption{\textbf{DefectNet workflow for predicting defect chemical elements and concentrations from phonon spectra.} Starting from a pristine parent material, doped supercells with substitutional dopants are generated and processed through high-throughput PDoS calculations using machine learning interatomic potential (MLIP). 
  An initial dopant guess, either manually provided or provided by a dopant recommender, is encoded along with the PDoS and parent material composition as input to DefectNet. It first employs multi-channel convolutions to extract spectral features, followed by a multi-head attention mechanism that integrates contextual relationships between defects and PDoS features. A final multilayer perceptron (MLP) outputs predicted defect chemical identities and concentrations, represented as an impurity-averaged configuration (IAC).
  }
  \label{fig1}
\end{figure}

Equipped with the reduced IAC defect representation (Eq.\,\ref{dopants}), we now introduce the workflow of the DefectNet model, shown in Fig.\,\ref{fig1}. The goal is to identify point-defect chemical species and concentrations directly from PDoS non-destructively. 
\textit{Data Generation:} We construct a dataset of 16,000 doped supercells derived from 2,000 pristine materials \cite{kim2020band}, covering a wide range of binary, ternary, quaternary, and quinary semiconductors (Fig.\,\ref{fig2}b). Substitutional defects are selected from the first 56 elements of the periodic table, excluding noble gases (Fig.\,\ref{fig2}a). These defects are introduced individually or in combination to reflect realistic co-doping scenarios, and may shed-light for future defect engineering in a high-dimensional design space. 
We employ a machine-learned recommender to guide substitutional element selection (more details in Methods) \cite{hautier2011data}. The recommender serves a dual purpose: it guides the training data generation, and when turned on, it guesses the potential defect species in a supercell.
We perform high-throughput structural relaxation and phonon calculations on both pristine crystals and doped supercells using the MACE-MP-0 foundational MLIP \cite{batatia2023foundation} under the frozen phonon formalism. We also apply Gaussian smearing to the simulated PDoS curves to mimic the experimental resolution function.  
\textit{Input for DefectNet:} The model accepts four inputs: the parent pristine crystal composition, the PDoS of the pristine crystal, the PDoS of the doped system, and an initial guess of likely defect chemical species. This guess can be provided manually by human intuition or prior knowledge, or generated automatically through a defect recommender, which is an ML-based probabilistic model to suggest the most likely substitutional elements \cite{hautier2011data}. 
While our training data are simulated, the framework is designed to work with experimental spectra such as those from INS through a fine-tuning process. 
\textit{Model architecture and defect attention:} Each input is processed through a separate channel. The composition and PDoS curves are fed through multi-channel convolution layers to generate a latent embedding $V$. The defect recommendation is embedded into a query vector $Q$, which is then combined with $V$ in a multi-head attention module \cite{vaswani2017attention}. This allows the model to focus on spectral features that are most informative for identifying specific defects, which is especially important at low concentrations or when multiple dopants are present. The resulting features are passed through multi-layer perceptrons (MLPs), and a final hard-masking step filters the predictions according to the candidate defects. 
\textit{Output:} Finally, the output features are hard-masked based on the initial defect guess, eliminating concentrations of dopants not included in the guess. This yields DefectNet’s predicted defect concentrations constrained to the set of initially assumed defect species. As a result, missed and incomplete guesses may prevent the model from recovering certain dopants. 
More details on the model architecture are discussed in Methods and Supplementary Information 2.

\subsection*{Defect signatures in vibrational spectra}
Before applying DefectNet, we first demonstrate that vibrational spectra can indeed encode defect profiles, which offers an intuitive insight of DefectNet.
Based on the quantum many-body theory of neutron scattering \cite{coleman2015introduction,fu2024anomalous}, the defective and pristine PDoS can be expressed in the following form under certain assumptions for analytical convenience (see Supplementary Information 1):
\begin{equation}
D(E)-D_0(E) = \int \lambda(\mathbf{q})\sum_{\alpha,\beta} n_{\alpha} n_{\beta} g_{\alpha\theta} g_{\beta\theta}\text{d}\mathbf{q} + \text{Defect-pristine cross terms}, 
\label{eq:retarded}
\end{equation}
where $D(E)$ is the doped PDoS, $D_0(E)$ represents the pristine PDoS, and $\lambda(\mathbf{q})$ is the prefactor depending on the microscopic properties with $\mathbf{q}$ dependence.

\begin{figure}
  \centering
  \includegraphics[width=0.9\textwidth]{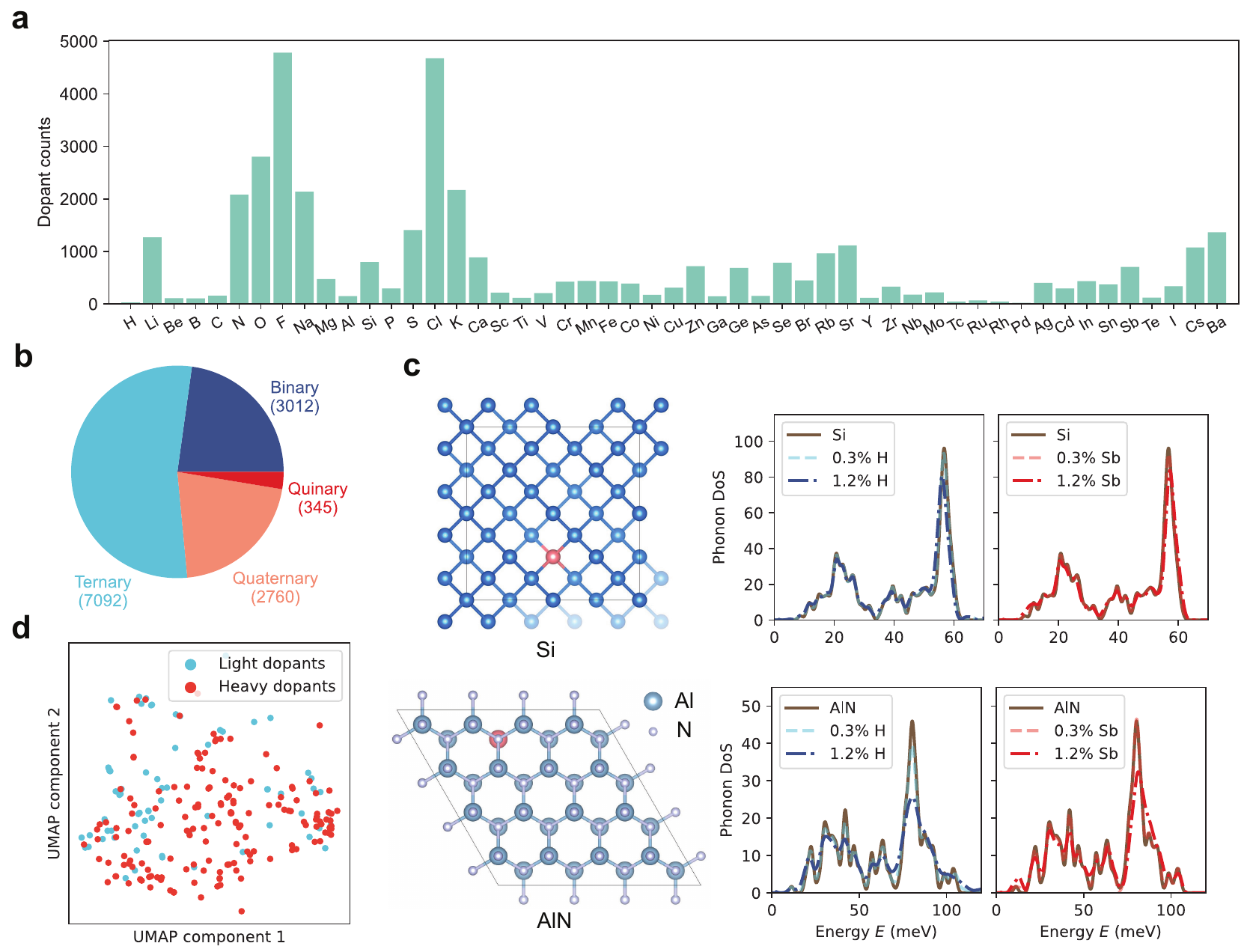}
  \caption{\textbf{Dataset composition and feature analysis for DefectNet.} \textbf{a.} Histogram of dopant elements across all generated doped supercells, indicating chemical diversity. \textbf{b.} Composition breakdown of parent compounds, ranging from binary to quinary systems, with ternary compounds most common. \textbf{c.} Phonon density-of-states (PDoS) comparisons between pristine and $\sim$ 0.3\%, 1.2\% doped systems for two prototypical semiconductors: Si and AlN, with spectral shifts and broadening peaks introduced by light (H) and heavy (Sb) dopants. \textbf{d.} UMAP dimensionality reduction of the PDoS dataset used for DefectNet training, where a subset of materials is colored according to defect mass (blue: lighter than Ne; red: heavier than Br). The UMAP projection reveals clustering based on defect mass, demonstrating a level of feature separability in PDoS space and supporting its suitability for machine learning.}
  \label{fig2}
\end{figure}

Eq.\,\ref{eq:retarded} establishes a direct but nonlinear relationship between the defect concentration and the PDoS under the given approximation, which poses a challenge for conventional methods to be deconvoluted to extract defect information.  
To demonstrate the feasibility of machine learning on defect features from PDoS, we analyze simulated spectra of doped crystals using spectral comparison and unsupervised learning (Fig.\,\ref{fig2}).  
Fig.\,\ref{fig2}c compares the PDoS of pristine, 0.3\%, and 1.2\% doped silicon (Si) and aluminum nitride (AlN).   
Lighter defects like hydrogen (H) introduce high-frequency modes, while heavier defects like antimony (Sb) cause redshifts. These subtle but systematic shifts demonstrate the feasibility of machine learning to resolve defect chemical species from vibrational spectra. The meV-level energy shift due to the defects lies within the feasibility range of the actual experiment of INS. We further apply Uniform Manifold Approximation and Projection (UMAP) \cite{mcinnes2018umap} for dimensionality reduction, which projects high-dimensional inputs into a lower-dimensional space while maintaining relative distances between nearby points, to the 16,000 spectra.  Fig.\,\ref{fig2}d shows the resulting 2D embeddings clustered by defect mass, further confirming that defect-specific vibrational signatures are encoded and exhibit partial separability, even in unsupervised settings. 

\subsection*{DefectNet for prediction of defect identify and concentration}
Now that we have established the learnability of defect signatures from PDoS with the IAC representation, we proceed to train DefectNet and demonstrate its generalization capabilities. 
We start by evaluating the performance of DefectNet on simulated PDoS for prototypical binary (SiC and AlAs) and ternary (AgGaS$_2$ and InCuSe$_2$) semiconductors.  

\begin{figure}
  \centering
  \includegraphics[width=0.9\textwidth]{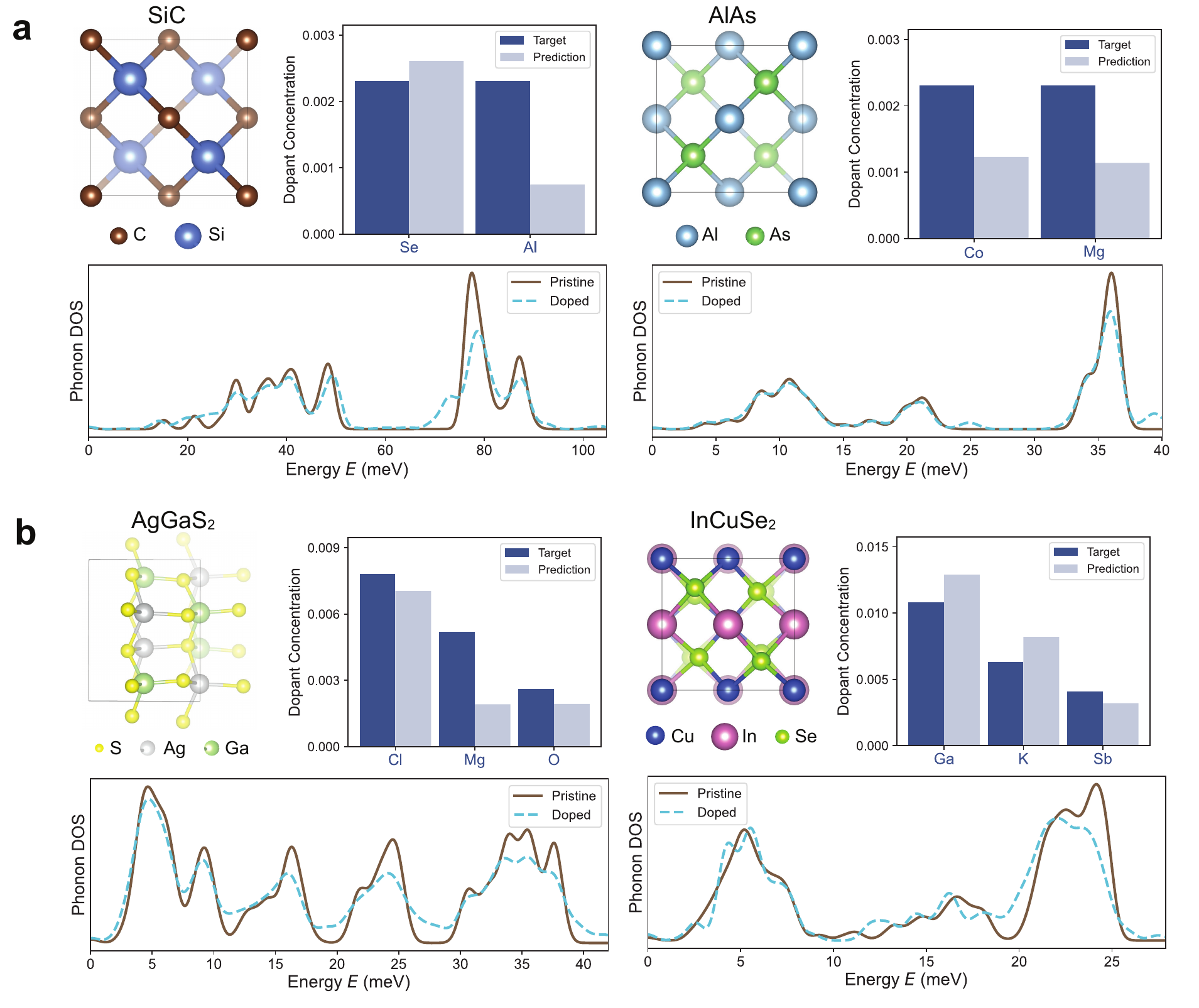}
  \caption{\textbf{Representative predictions of DefectNet on prototypical semiconductors.} \textbf{a.} Results for binary semiconductors SiC and AlAs; \textbf{b.} Results for ternary semiconductors AgGaS$_2$ and InCuSe$_2$. 
  For each material, we show the atomic structure of the parent crystal (top left), a bar plot comparing predicted and true defect concentrations (top right), and the target phonon density-of-states (PDoS) before and after doping (bottom).}
  \label{fig3}
\end{figure}

SiC and AlAs are semiconductors with large bandgaps valued for high-power electronics and heterostructure design, respectively \cite{hudgins2003wide,ruff1994sic,paveliev2012experimental}. Fig.\,\ref{fig3}a shows the PDoS of pristine and doped configurations alongside predicted and true dopant concentrations. Despite low doping levels ($\sim$ 1\%), DefectNet largely captures subtle vibrational shifts and reliably recovers the correct dopant concentrations. 
We next apply DefectNet to more chemically complex ternary semiconductors AgGaS$_2$ and InCuSe$_2$ (Fig.\,\ref{fig3}b). AgGaS$_2$ is used in infrared nonlinear optics \cite{huang2022density,zhang2025theoretical}, while InCuSe$_2$ is a promising material in thin-film photovoltaics \cite{zhang1998defect,ali2025highly}. These materials contain multiple inequivalent atomic sites and more diverse vibrational modes. Still, DefectNet tracks PDoS changes and is capable of inferring dopant concentrations, showing its robustness across a broad range of structural and chemical complexities.

\begin{figure}
  \centering
  \includegraphics[width=\textwidth]{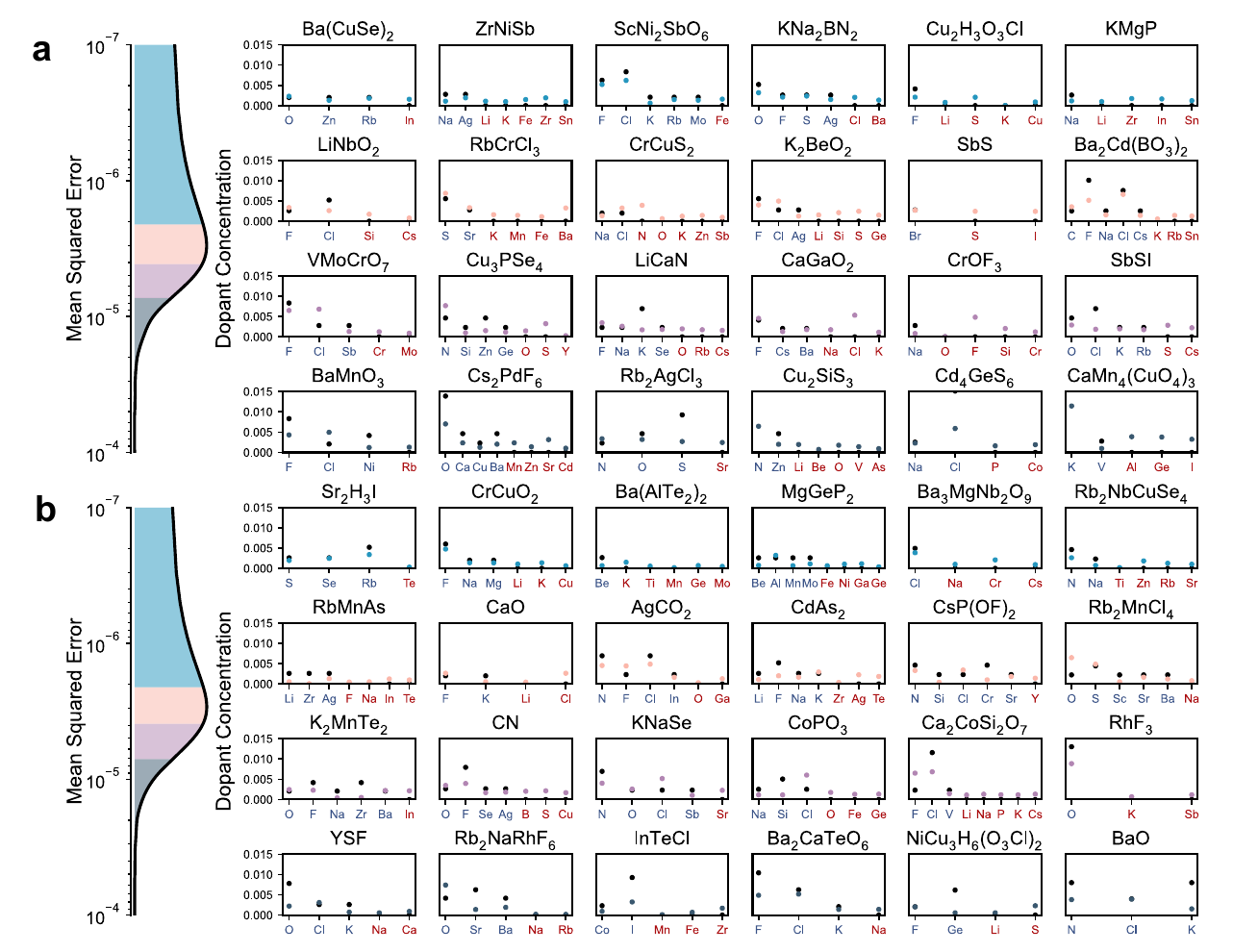}
  \caption{\textbf{Evaluating DefectNet predictions on diverse materials and defect configurations.} \textbf{a.} In-distribution, known-parent-material, unknown-defect predictions, where training and testing sets are generated by randomly splitting phonon density-of-states (PDoS) curves from the same set of pristine parent materials. \textbf{b.} Out-of-distribution, unseen-parent-material, unknown-defect predictions, where the entire pristine parent materials are excluded from training and used only for testing. In both cases, each panel compares the predicted (colored dots) and ground-truth (black dots) defect concentrations with multiple coexisting defects. Red colored chemical elements indicate distractor species introduced only during testing to evaluate model robustness. 
  The leftmost plots display the distribution of mean squared error (MSE) in predicted defect concentrations, grouped into quartiles based on overall model performance sorted by error (top to bottom: low to high).}
  \label{fig4}
\end{figure}

To further evaluate DefectNet’s generalizability, we test its performance on the full defect dataset containing multiple coexisting dopants, including ``distractor'' defects that appear in the initial guess but do not actually exist in the ground truth. This setting evaluates the robustness of DefectNet to detect true dopants while suppressing false-positive candidates. Fig.\,\ref{fig4} summarizes these results by comparing predicted defect concentrations (colored dots) and ground-truth values (black dots), grouped by quartiles of prediction mean-square error (MSE). In the in-distribution case (Fig.\,\ref{fig4}a), where pristine parent materials are seen in training data but their defect information is unknown, DefectNet achieves high-fidelity predictions across a wide range of defect chemical species and concentrations. Even with distractor defects included in the input, DefectNet can eliminate the distractors and identify the true dopants. Fig.\,\ref{fig4}b shows the more demanding out-of-distribution case, where even pristine materials are not seen during training. Although the prediction accuracy is slightly reduced, DefectNet continues to capture the key dopant signatures and assigns near-zero concentration values to most distractors, reflecting strong model generalizability. Additional tests are shown in Supplementary Information 3.

While DefectNet performs well on synthetic PDoS data generated by MLIPs, these spectra differ from those obtained via DFT or experiments. Discrepancies in force-field accuracy, measurement noise, and uncontrolled defect environments can shift the data distribution, making defect signatures especially subtle and hard to resolve in cases such as lightly doped semiconductors. To effectively apply DefectNet to real experimental data, additional steps, such as fine-tuning on the target dataset, are necessary to bridge this domain gap and ensure direct applicability to experimental data.

\subsection*{Fine-tuning DefectNet on experimental data}
\begin{figure}
  \centering
\includegraphics[width=1.0\textwidth]{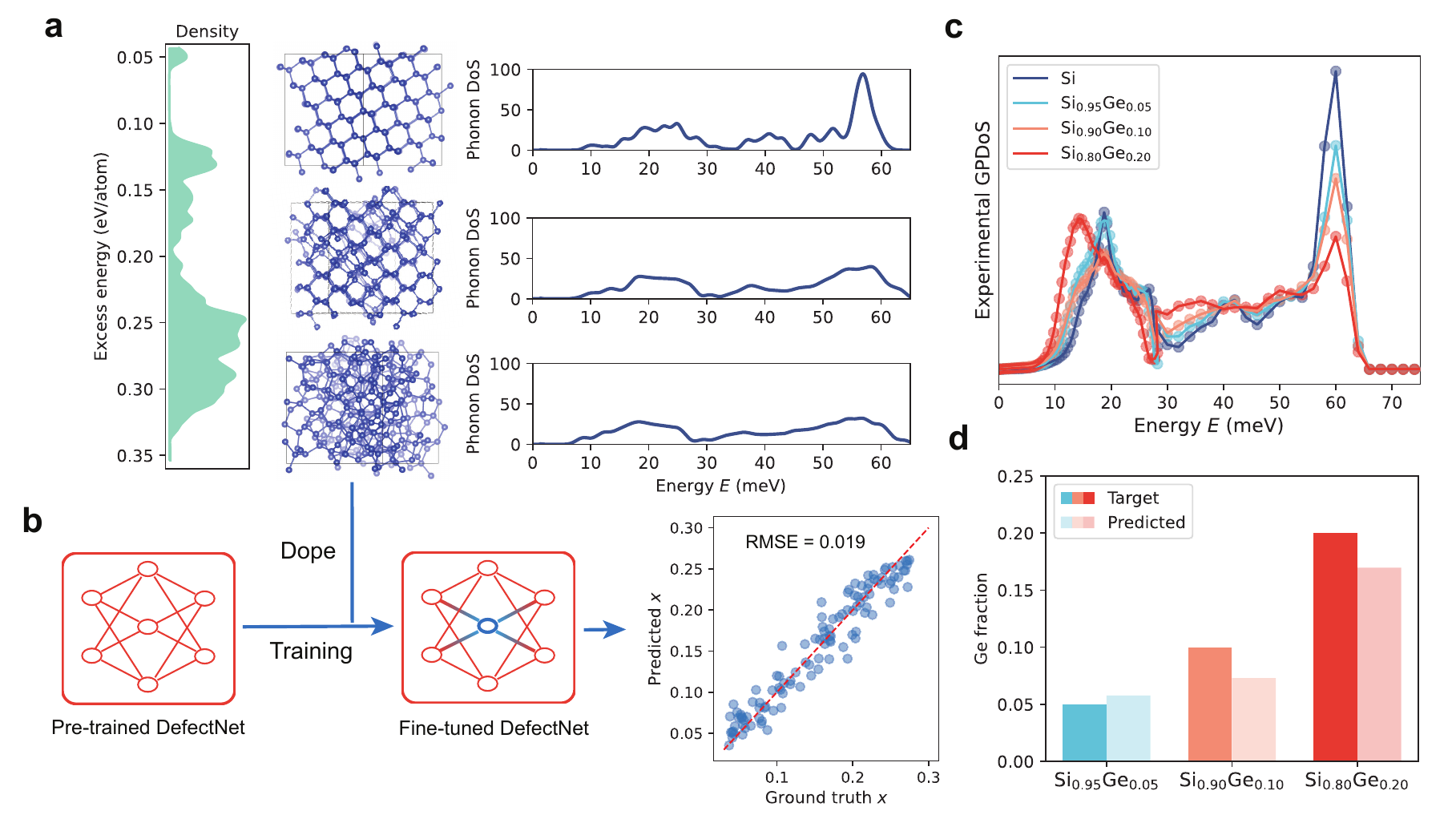}
  \caption{\textbf{Fine-tuning DefectNet for experimental validation on SiGe alloy.} \textbf{a.} Distribution of excess potential energies for disordered Si structures, indicating varied degrees of disorder. Three representative atomic configurations are shown alongside their computed PDoS, ranging from near-crystalline (top) to highly amorphous (bottom).
  \textbf{b.} Workflow of fine-tuning DefectNet: a pre-trained DefectNet (red box) is adapted to Si$_{1-x}$Ge$_x$ through transfer learning to obtain the fine-tuned DefectNet with updated weights (red box with blue fine-tuned weights). The scatter plot compares predicted and truth Ge fraction $x$ with a low RMSE=0.019. 
  \textbf{c.} Experimental generalized phonon density-of-states (GPDoS) for pure Si and Si$_{1-x}$Ge$_{x}$ alloys with $x=0.05$, $0.10$, and $0.20$, extracted from Ref.\,\cite{dhital2012inelastic}. \textbf{d.} Comparison of predicted Ge fractions $x$ from fine-tuned DefectNet against experimentally reported values for the same set of samples.
  }
  \label{fig5}
\end{figure}

To establish the practical utility of DefectNet, we fine-tune and test it on experimental data. 
In particular, we test DefectNet on thermoelectric SiGe alloy \cite{dhital2012inelastic} as well as MgB$_2$ superconductor. 
Here, we highlight the case of SiGe, since it poses additional challenge due to the structural disorders even in the undoped system. DefectNet addresses this challenge by relying only on the vibrational spectra and chemical formula, making it suitable to study non-crystalline systems and thereby significantly expands its applicability. We construct a training dataset of 100 amorphous Si supercells sampled from the Si-GAP-18 database \cite{bartok2018machine} using quenching simulations. These supercells span a wide range of disorders, from low-energy, near-crystalline states to highly disordered configurations. As the disorder increases, the PDoS broadens and the characteristic 60 meV optical phonon peak becomes suppressed (Fig.~\ref{fig5}a), providing parent materials for model fine-tuning. Subsequently, we simulate SiGe alloys by randomly substituting Ge into amorphous Si supercells, spanning various levels of disorder, at concentrations ranging from 0\% to 25\%.
Given access to an existing database of amorphous Si, this approach is more efficient than direct quenching of an alloy system while still capturing meaningful spectral variations across a broad structural space.
Vibrational properties are computed using MatterSim, a high-accuracy MLIP \cite{yang2024mattersim}. To match experimental observables, we compute the generalized phonon density-of-states (GPDoS), which represents a neutron-weighted average over atomic species:  
\begin{equation}
    \text{GPDoS}(E) \propto \sum_j g_j(E) \eta_j \frac{\sigma_j}{m_j}.
\end{equation}
where $g_j(E)$ is the partial PDoS for atom $j, \eta_j$ is the atomic fraction, $\sigma_j$ is the neutron scattering cross-section\cite{sears1992neutron}, and $m_j$ is the atomic mass. This results in 500 unique GPDoS data used to fine-tune DefectNet. 

After fine-tuning, DefectNet achieves a test set root squared mean squared error (RMSE) of 0.019, indicating strong predictive performance (Fig.\,\ref{fig5}b). We then apply the fine-tuned DefectNet model to experimental GPDoS data for Si$_{1-x}$Ge$_x$ alloy\cite{dhital2012inelastic}, with nominal alloying ratio of $x=$ 5\%, 10\%, and 20\% (Fig.\,\ref{fig5}c). DefectNet predicts Ge concentrations of 7\%, 13\%, and 22\%, respectively, which closely track the experimental trends (Fig.\,\ref{fig5}d). Given the inherent difficulty in precisely quantifying defects in amorphous materials, our results demonstrate strong predictive power of DefectNet when applying to experimental data. For the aluminum (Al)-doped multiband superconductor MgB$_2$ \cite{yokoo2004evidence}, the fine-tuned DefectNet reproduces the experimental trend of dopant concentration up to 25\%. More details are discussed in Supplementary Information 4.

\section*{Discussion}
In this work, we present DefectNet, a foundational machine learning model that enables non-destructive identification and quantification of point defects directly from vibrational spectra. Capable of resolving up to six coexisting substitutional defect species at concentrations as low as 0.2\%, DefectNet operates on PDoS data without requiring detailed atomic structures. The model is built upon a physically motivated IAC representation and enhanced with a spectral attention mechanism that boosts sensitivity to subtle vibrational shifts. When fine-tuned on experimental INS data, DefectNet can recover dopant concentrations in both crystalline and amorphous systems, as demonstrated in SiGe alloys and MgB$_2$ superconductors. These results not only establish the transferability of DefectNet to real-world materials, but also position INS as a powerful and element-sensitive platform for non-destructive defect reconstruction.

While promising, several challenges remain. First, the model’s sensitivity decreases at very low defect concentrations, where vibrational signatures become subtle and hard to distinguish from noise. Second, the current version is limited to substitutional doping; extending DefectNet to handle multiple point defect types, such as interstitials, vacancies, Frenkel pairs, or defect clusters, would significantly expand its scope. Third, despite strong generalization from simulated data, fine-tuning on experimental datasets remains essential. A model directly and robustly applicable to raw experimental spectra without retraining remains an aspirational but formidable goal. Finally, while INS offers exceptional resolution and chemical sensitivity, its limited accessibility constrains widespread deployment. Building a parallel DefectNet variant for more accessible techniques such as Raman spectroscopy would greatly enhance usability and integration in diverse laboratory environments.

Looking forward, DefectNet represents a step toward a unified, data-driven paradigm in defect science. Its architecture is naturally compatible with multimodal spectroscopic inputs and opens the door to inverse design of materials with targeted defect profiles. By combining physically informed representations, high-throughput simulation, scalable learning, and experimental fine-tuning, DefectNet charts a path toward automated, interpretable, and non-destructive defect engineering across the full complexity of real materials.

\section*{Methods}
\subsection*{Dopant recommender for DefectNet}
To determine viable dopant substitutions for each semiconductor, we employ a two-stage recommendation system. For each parent compound, we first consult the probabilistic substitutional doping model proposed by Hautier et al. \cite{hautier2011data}, which ranks element pairs by chemical compatibility and historical substitution trends. 
In cases where the Hautier model failed, often due to invalid oxidation states, we apply a fallback strategy based on Shannon ionic radius matching \cite{shannon1969effective,shannon1976revised} to identify plausible dopants. Only dopants that passed either filter are used in subsequent sampling. This recommendation framework ensures physically reasonable, chemically consistent dopant-host combinations across all DefectNet training data. Both dopant recommenders are implemented in the Pymatgen interface \cite{ong2013python}.

\subsection*{Dataset generation for universal DefectNet}
To generate a high-quality dataset for training the universal DefectNet, we randomly select 2,000 semiconductors from an established semiconductor database \cite{kim2020band}. Each structure is processed through the dopant recommender to identify $n$-type and $p$-type doping candidates. 
The host crystals are subsequently expanded into supercells containing between 433 and 500 atoms, with the size adaptively adjusted to ensure dilute doping concentrations with a lower bound of $\sim$ 0.2\%. Dopants are substituted into the host lattice, and each doped structure is relaxed, until all atomic forces converged below 0.01 eV/\AA.
Following structural relaxation, vibrational properties are evaluated using the finite-displacement method to compute PDoS, forming the spectral inputs to DefectNet.
All calculations for this pre-training step are performed using the MACE-MP0 MLIP \cite{batatia2023foundation} on the Atomic Simulation Environment (ASE) \cite{larsen2017atomic} and Pymatgen \cite{ong2013python} interfaces.

Additionally, we note that DefectNet imposes no strict constraint on the source of the vibrational spectroscopy. In practice, the doped sample’s spectrum is tractable from measurements like INS as a real-world input. 
For the pristine counterpart, users are free to choose the most credible or convenient source available: this may be first-principle DFT calculations, a machine-learned surrogate model \cite{chen2021direct,kong2022density}, a machine learning interatomic potential (MLIP), or even experimental data collected under the same lab conditions as the doped sample to ensure maximal consistency.

\subsection*{Methods of machine learning}
We use scikit-learn \cite{scikit-learn} and UMAP \cite{mcinnes2018umap} package for dimensional reduction on PDoS data.
The PyTorch \cite{paszke2019pytorch} package is used for DefectNet training. 
DefectNet is a hybrid convolutional-attention neural network predicting dopant species and concentrations from vibrational spectra. The spectral features are first processed by two blocks of multi-channel 1D convolutional layers, each followed by a ReLU activation. The convolved representation and initial dopant guess vector are then projected to value ($V$) and query ($Q$) embeddings and passed through a multi-head self-attention module.
The attention mechanism follows the standard scaled dot-product self-attention formulation \cite{vaswani2017attention}:
\begin{equation}
\mathrm{Attention}(Q, K, V) = \mathrm{softmax}\left( \frac{QK^\top}{\sqrt{d_k}} \right) V,
\end{equation}
where $Q$, $K$, and $V$ are the query, key, and value matrices, respectively, and $d_k$ is the dimensionality of the key vectors. In the case of self-attention used in DefectNet, we have $K = V$ as both are derived from the same input.

To enhance representational capacity, DefectNet employs multi-head self-attention:
\begin{equation}
\mathrm{MultiHead}(Q, V, V) = \mathrm{Concat}(h_1, h_2, \ldots, h_H) W^O,
\end{equation}
where each attention head is computed as:
\begin{equation}
h_i = \mathrm{Attention}(QW_i^Q, VW_i^K, VW_i^V),
\end{equation}
and $W_i^Q$, $W_i^K$, $W_i^V$, and $W^O$ are learnable weight matrices specific to each head. Hard masking is applied to suppress attention weights for zero dopant concentrations, ensuring that attention is only computed over the user's initial guessed dopants.

\section*{Data and code availability}
The data used in this study are numerically generated using our code implementing MLIP, and the code used in this study is available at \url{https://github.com/vipandyc/MLforDefect}.

\section*{Acknowledgments}
The authors thank KA Nelson and N Andrejevic for helpful discussions, and thank S Wilson for providing the experimental data on SiGe alloy, originally published in Ref.\,\cite{dhital2012inelastic}.
M.C. acknowledges the support of U.S. Department of Energy (DOE), Office of Science (SC), Basic Energy Sciences (BES), award No. DE-SC0021940. C.F. acknowledges support from DOE BES award No. DE-SC0020148. B.Y. National Science Foundation (NSF) Designing Materials to Revolutionize and Engineer our Future (DMREF) Program with award No. DMR-2118448. A.C. acknowledges support from NSF ITE-2345084. D.L.A. and Y.C. were supported by the Scientific User Facilities Division, Office of Basic Energy Sciences, U.S. Department of Energy, under Contract No. DE-AC0500OR22725 with UT Battelle, LLC.
A portion of computational simulation results were obtained using the Lonestar6 computing system at the Texas Advanced Computing Center. M.L. acknowledges the support from the Class of 1947 Career Development Chair and the support from R. Wachnik.

\bibliography{refs.bib} % Tell bibtex which .bib file to use

\end{document}

\maketitle
\flushbottom
\tableofcontents
\thispagestyle{empty}

\section{Defect Representation Physics}
\subsection{Defect representation with impurity average}
We start by introducing our theoretical framework of the defect representation. To represent crystalline systems with defects, we begin by defining a generalized atomic density field $\boldsymbol{\rho}(\mathbf{r})$ in real space for an ideal, pristine crystal:
\begin{equation}
    \mathbf{\rho}_{\text{pristine}}(\mathbf{r}) = \sum_{j} \vec{f}_{j} \delta(\mathbf{r}-\mathbf{R}_j),
\end{equation}
where $j$ runs over all atoms in the material, and $\mathbf{R}_j$ is the corresponding spatial position. $\vec{f}_{j}$ is constructed as the atomic feature vector, which could involve multiple features $\vec{f}_{j} = \left(1,Z_j,m_j,s_j,r_j,...\right)$ where 1, $Z_j$, $m_j$,$S_j$, and $r_j$, are one-hot, charge, mass, spin, and radius features at the specific positions and finally can be expressed as:
\begin{equation}
    f_{j\theta} = \text{Embed}_\theta (1,Z_j,m_j,s_j,r_j,...)
\end{equation}
which make all input physical features of each atom $j$ optimized for their embedding mixture by an ensemble embedding layer $\text{Embed}_\theta$. Suppose we limit ourselves to thinking only about the local point-like perturbation introduced by the defects.  We model defects as perturbations of these feature vectors or as contributions to off-lattice positions, thus
\begin{equation}
    \boldsymbol{\rho}(\mathbf{r}) = \sum_j \vec{f}_j\, \delta(\mathbf{r} - \mathbf{R}_j) + \sum_{j\alpha} \vec{g}_{j\alpha }\, \delta(\mathbf{r} - \mathbf{R}_{j\alpha}),
\end{equation}
where $\vec{g}_{j\alpha}$ denotes the feature vector of a defect of type $\alpha$ at location $\mathbf{R}_{j\alpha}$, with same embedding types as $\vec{f}_j$ represented before. To reduce the complexity under the thermodynamic limit with an infinite number of lattice sites/atoms and change from the linear vector to embedding, we shift to the momentum space with the embedding form:
\begin{equation}
    \boldsymbol{\rho}(\mathbf{q}) = \boldsymbol{\rho}_{\mathrm{pristine}}(\mathbf{q}) + \boldsymbol{\rho}_{\mathrm{defect}}(\mathbf{q}) = \sum_j f_{j\theta}\, e^{i\mathbf{q} \cdot \mathbf{R}_{j}} +  \sum_{j\alpha} g_{j\alpha\theta}\, e^{i\mathbf{q} \cdot \mathbf{R}_{j\alpha}}.
    \label{rhoq}
\end{equation}
Now we introduce the impurity average. For an impurity-position-dependent physical quantity $F(\{\mathbf{R}_j\})$ and its thermodynamic ensemble average $\langle F(\{\mathbf{R}_j\}) \rangle$, the impurity average can be carried out by a joint probability $P(\mathbf{R}_1,\mathbf{R}_2,\cdots,\mathbf{R}_{N_D}) \equiv P(\{\mathbf{R}_j\})$ that performs weighted average of the impurity spatial distribution,
 \begin{equation}
     \overline{\langle F \rangle} = \int \prod_{j} d^3\mathbf{R}_j P(\{\mathbf{R}_j\}) \langle F(\{\mathbf{R}_j\})\rangle\,.
     \label{eq:impurity_average_si}
 \end{equation}
In particular, assuming that the positions of each impurity are distributed independently and each impurity is distributed uniformly inside the solid, then we have 
\begin{equation}
P(\mathbf{R}_1,\mathbf{R}_2,\cdots,\mathbf{R}_{N_D})=\prod_{j=1}^{N_D} P(\mathbf{R}_j)=\frac{1}{\mathcal{V}^{N_D}}\,.
\label{iid}
\end{equation}
where $\mathcal{V}$ is the system volume and $N_D$ is the number of defects assuming one type of defects exist. 
Notice that the probability here is assumed to be uniform, but can be generalized in other cases. The first equality comes from the assumption of independence between pairs of defects. The second equality comes from the uniformity of defect spatial distribution.

While the atomic density with defect depends on the specific position of the impurities, we can perform the impurity average with the thermodynamic ensemble average $\langle \sum_{j\alpha} \vec{g}_{j\alpha}\, e^{i\mathbf{q} \cdot \mathbf{R}_{j\alpha}}\rangle$ with the joint spatial probability of the defect configuration $P(\{\mathbf{R}_{j{\alpha}}\})$, which finally motivates us to define our Impurity-Averaged
Configuration (IAC) for defects:
\begin{equation}
\overline{\left\langle\boldsymbol{\rho}_{\text {defect }}(\mathbf{q})\right\rangle}=\bigotimes_{j \alpha}\mathbb{E}_{\left\{\mathbf{R}_{j \alpha}\right\} \sim P_\phi}\left[g_{j\alpha\theta} \times e^{i \mathbf{q} \cdot \mathbf{R}_{j \alpha}}\right] .
     \label{eq:impurity_average}
 \end{equation}
Here $\bigotimes_{j \alpha}$ denotes a tensor product,  $\times$ represents elemental multiplication, and $\mathbb{E}$ is the expectation value. The probability model $P_\phi$ contains all the spatial distribution relations of defects. This can be explicitly parametrized via pairwise correlation functions, learned via deep neural networks or sampled from known experimental distributions.

With the IAC defined above, one can readily embed the general defect-related information through averaging all the detailed positional information, given the appropriate probability distribution $P$. Since the feature embedding $g_{j\alpha\theta}$ is introduced in general, the IAC can be applied to represent multiple configurations of correlated defects, including positional, charge, and spin correlations, which can even be further connected to general observables. 
Assuming the defects are distributed randomly and independently, the IAC representation reduces to the lowest order, which is only dependent on defect concentrations:
\begin{align}
    \boldsymbol{\rho}_{\rm{IAC}}(\mathbf{q}) = \boldsymbol{\rho}(\mathbf{q}) - \boldsymbol{\rho}_{\mathrm{pristine}}(\mathbf{q}) = \bigoplus_\alpha n_{\alpha} \langle g_{j\alpha\theta} \rangle_j\, \delta_{\mathbf{q}0}= \bigoplus_\alpha n_{\alpha} g_{\alpha\theta}\, \delta_{\mathbf{q}0} 
\label{dopants}
\end{align}
where $n_{\alpha}$ is the defect density of type-$\alpha$ defects normalized from the number of defects and $g_{j\alpha\theta}$ is the expectation value of the feature embedding for that defect type $\alpha$. 
Now if we focus on the first trivial atomic feature of each feature embedding $g_{j\alpha\theta}$, the resulting IAC descriptor turns out to be a concatenated vector $\boldsymbol{\rho}_{\rm{IAC}}(\mathbf{q}) = (...,n_{\alpha},...)^T \delta_{\mathbf{q} 0}$, which gives us the dopant concentration of each defect $\alpha$ in a given material.

\subsection{Inelastic scattering for the phonon dispersion}
To link the vibrational spectra and the IAC representation, we further define the retarded density-density response function and adapt the inelastic neutron scattering for the nuclear scattering to see how the phonon density of state can be considered as one of the features for representing the defects. The derivation is based on and adapted to the neutron inelastic scattering derivation \cite{bruus2004many,coleman2015introduction,fu2024anomalous}, but with the defects in the signal. The neutron inelastic scattering is utilized to probe the nuclear density-density response function to understand the phonon dispersion.
The nuclear density-density response function ${\chi _{\rho \rho }}({\bf{q}},i{\omega _n})$ can be expressed as, 
\begin{equation}
\begin{split}
{\chi _{\rho \rho }}({\bf{q}},i{\omega _n})& = \int\limits_0^\beta  {\chi ({\bf{q}},\tau  \equiv {\tau _1} - {\tau _2}){e^{i{\omega _n}\tau }}d\tau } \\
{\chi _{\rho \rho }}({\bf{q}},{\tau _1} - {\tau _2}) &=  - {{\rm{T}}_\tau }{\left\langle {\left( {\rho_{n} ({\bf{q}}{\tau _1}) - \left\langle {\rho_{n} ({\bf{q}}{\tau _1})} \right\rangle } \right)\left( {\rho_{n} ( - {\bf{q}},{\tau _2}) - \left\langle {\rho_{n} ( - {\bf{q}},{\tau _2})} \right\rangle } \right)} \right\rangle _H}\\
 &= \frac{1}{\beta }\sum\limits_n {\chi ({\bf{q}},i{\omega _n}){e^{ - i{\omega _n}({\tau _1} - {\tau _2})}}} 
\end{split}
\end{equation}
in which ${\chi _{\rho \rho }}({\mathbf{q}},{\tau _1} - {\tau _2})$ is the density-density response function in “interaction picture” in imaginary time $\tau$, where the interaction picture means the average is ${\left\langle {} \right\rangle _H}$, referring to the full materials Hamiltonian H. Here ${\omega _n} = \frac{{2\pi n}}{\beta },n = 0, \pm 1, \pm 2,...$ is Bosonic Matsubara frequency, while ${{\rm{T}}_\tau }$ denotes the imaginary-time ordering operator. $\beta =1/k_{\rm{B}}T$ is the inverse temperature, while $k_{\rm{B}}$ is the Boltzmann constant. $\rho_{n}$ represents the nuclear density function. We expand the density function in the lowest order with the embedded defect in a similar embedding form:
\begin{equation}
\begin{split}
    {\hat{\rho}_{n}}(\mathbf{q}{\tau_1}) \xrightarrow{\text{Embedding}} \sum\nolimits_{l}{f}_{l\theta}  e^{i \mathbf{q}\cdot \mathbf{R}_l^0}\left( {1 + i{\bf{q}} \cdot {{{\bf{\hat u}}}_l}({\tau _1})} \right) + \sum\nolimits_{j{\alpha}}{g}_{j{\alpha \theta}}e^{i\mathbf{q}\cdot \mathbf{R}^0_{j{\alpha}}}\left( {1 + i{\bf{q}} \cdot {{{\bf{\hat u}}}_{j{\alpha}}}({\tau _1})} \right)
\end{split}
\end{equation}
Here ${{{\bf{\hat u}}}_l}({\tau _1})$ is the displacement for the pristine atomic density. ${{{\bf{\hat u}}}_{j\alpha}}({\tau _1})$ is the displacement of the $j$th defect with $\alpha$ type. All the “hat” operators are in interaction picture w.r.t. the non-interacting part of the materials Hamiltonian, i.e. $\hat A(\tau ) = {e^{ + \tau {H_0}}}A{e^{ - \tau {H_0}}}$. This embedding and approximation means we treat the effect of defects or embedded inhomogeneities at the operator level, by modifying the expression of the embedded density operator to account for deviations from the ideal crystal structure. However, we assume that these modifications do not alter the underlying lattice dynamics. Therefore, all correlation functions are assumed to be evaluated with respect to the non-interacting Hamiltonian $H_0$. As a result, when we consider the embedded density correlation, it will have three correlation contributions, the pristine crystal's contribution $\chi_{PP} $, pristine-defect interaction contribution $ \chi_{PD}$, defect-defect contribution $\chi_{DD}$, while 
\begin{equation}
    \begin{split}
    & \chi_{\rho\rho} \xrightarrow{\text{Embedding}} \chi_{PP} + \chi_{PP} + \chi_{DD} \\
        &\chi_{PP} = \sum\nolimits_{l,l' = 1}^N {{f}_{l \theta}  e^{-i \mathbf{q}\cdot \mathbf{R}_l^0}{f}_{l'\theta}  e^{i \mathbf{q}\cdot \mathbf{R}_l'^0}{{\mathrm{T}}_\tau }{{\left\langle {( - i){\bf{q}} \cdot {{{\bf{\hat u}}}_l}({\tau _1})i{\bf{q}} \cdot {{{\bf{\hat u}}}_{l'}}({\tau _2})} \right\rangle }_{H_0}}} \\
        &\chi_{PD} = 2\sum\nolimits_{l,j{\alpha}} {{f}_{l\theta}  e^{-i \mathbf{q}\cdot \mathbf{R}_l^0}{g}_{j{\alpha \theta}}  e^{i \mathbf{q}\cdot \mathbf{R}_{j{\alpha}}^0}{{\mathrm{T}}_\tau }{{\left\langle {( - i){\bf{q}} \cdot {{{\bf{\hat u}}}_l}({\tau _1})i{\bf{q}} \cdot {{{\bf{\hat u}}}_{j{\alpha}}}({\tau _2})} \right\rangle }_{H_0}}}\\
        &\chi_{DD} = \sum\nolimits_{j{\alpha},j'{\alpha'}} {{g}_{j{\alpha}\theta}  e^{-i \mathbf{q}\cdot \mathbf{R}_{j{\alpha}}^0}{g}_{j'{\alpha}'\theta}  e^{i \mathbf{q}\cdot \mathbf{R}_{j'{\alpha}'}^0}{{\mathrm{T}}_\tau }{{\left\langle { (-i){\bf{q}} \cdot {{{\bf{\hat u}}}_{j{\alpha}}}({\tau _1}) i{\bf{q}} \cdot {{{\bf{\hat u}}}_{j'{\alpha}'}}({\tau _2})} \right\rangle }_{H_0}}}
    \end{split}
\end{equation}
For $\chi_{PP}$, it generates an embedding of the density response function for the pristine part, and the connection between this response function and the inelastic neutron scattering is standard now \cite{coleman2015introduction}. To simplify the discussion, we take the low-temperature limit to ignore the Debye-Waller factor. We have the following equation
\begin{equation}
    ( - i){\bf{q}} \cdot {{{\bf{\hat u}}}_l}(\tau_1) = \sum\limits_{{\bf{q''}}} {\sqrt {\frac{\hslash }{{2MN{\omega _{{\bf{q''}}}}}}} \left( {a_{{\bf{q''}}}^{}({\tau _1}) + a_{ - {\bf{q''}}}^ + ({\tau _1})} \right){e^{i{\bf{q''}} \cdot {\bf{R}}_l^0}}\left( { - i{\bf{q}} \cdot {\varepsilon _{{\bf{q''}}}}} \right)}
\end{equation}
For $\chi_{PD}$ and $\chi_{DD}$, cases are a little bit different, as we require the impurity average to conduct the averaging among the whole summation. But what we need to deal with is the momentum in exponential form from the density expansion and the displacement operator. So we may have the following results to simplify the expression
\begin{equation}
\sum_{\mathbf{q''}}\sum_{j{\alpha}}{g}_{j{\alpha}\theta}e^{i\mathbf{q}\cdot \mathbf{R}^0_{j{\alpha}}}e^{i \mathbf{q''}\cdot \mathbf{R}_{j{\alpha}}^{0}} = \sum_{\mathbf{q''}}\sum_{j{\alpha}}{g}_{j{\alpha}\theta}e^{i(\mathbf{q}+\mathbf{q}'')\cdot \mathbf{R}^0_{j{\alpha}}} =\sum_{\mathbf{q''}\alpha} {{g}_{{\alpha}\theta}}n_{{\alpha}}\delta_{\mathbf{q}(-\mathbf{q}'')}
\end{equation}
The $\mathbf{q}''$ term comes from the expansion of the displacement operator, and we have placed the impurity average with the independent uniform distribution in the final equality. Here we have assumed the corresponding displacement operator on the defect is expanded in a similar perfect crystal's plane-wave basis, but with "distorted" mass, frequency, and (impurity) positions, and then we have 
\begin{equation}
    ( - i){\bf{q}} \cdot {{{\bf{\hat u}}}_{j\alpha}}(\tau_1) = \sum\limits_{{\bf{q''}}} {\sqrt {\frac{\hslash }{{2M_{j\alpha}N{\omega _{j\alpha,{\bf{q''}}}}}}} \left( {a_{{\bf{q''}}}^{}({\tau _1}) + a_{ - {\bf{q''}}}^ + ({\tau _1})} \right){e^{i{\bf{q''}} \cdot {\bf{R}}_{j\alpha}^0}}\left( { - i{\bf{q}} \cdot {\varepsilon _{{\bf{q''}}}}} \right)}
\end{equation}
While this is an approximation—since defect modes may deviate from bulk phonons—the leading effects are captured by using defect-specific mass and effective frequencies, and finally by performing an ensemble average over the defect positions. This yields a defect-sensitive yet analytically tractable phonon response framework. Conducting such a calculation will lead to three channels with different defect number densities and the averaged feature factors as the contribution. Then for $\chi_{PD}$ and $\chi_{DD}$, we have
\begin{equation}
\begin{split}
    \chi _{PD}^{}({\bf{q}},{\tau _1} - {\tau _2}) &= \sum_{\alpha }\frac{{\hslash N{f}_{\theta} n_{{\alpha}}{{g_{{\alpha \theta}}}}}}{{2\sqrt{MM_{\alpha, Davg}}\sqrt {{\omega _{\bf{q}}}{\omega _{ \alpha, Davg,- {\bf{q}}}}} }}\left( {{\bf{q}} \cdot {\varepsilon _{\bf{q}}}} \right)\left( {{\bf{q}} \cdot {\varepsilon _{ - {\bf{q}}}}} \right)D_{\bf{q}}^{(0)}({\tau _1} - {\tau _2}) \propto \sum_{\alpha }n_{{\alpha}}{{g_{{\alpha \theta}}}}\\
    \chi _{DD}^{}({\bf{q}},{\tau _1} - {\tau _2}) &=  \sum_{\alpha \beta}\frac{{\hslash n_{{\alpha}}n_{{\beta}}{{g_{{\alpha\theta}}g_{{\beta\theta}}}}}}{{2\sqrt{M_{\alpha, Davg}M_{\beta,Davg}}\sqrt {{\omega _{\alpha, Davg,\bf{q}}}{\omega _{\beta,Davg,- {\bf{q}}}}} }}\left( {{\bf{q}} \cdot {\varepsilon _{\bf{q}}}} \right)\left( {{\bf{q}} \cdot {\varepsilon _{ - {\bf{q}}}}} \right)D_{\bf{q}}^{(0)}({\tau _1} - {\tau _2})\propto \sum_{\alpha \beta}n_{{\alpha}}n_{{\beta}}{{g_{{\alpha\theta}}g_{{\beta\theta}}}}
\end{split}
\end{equation}
in which $D_{\mathbf{q}}^{(0)}\left(\tau_1-\tau_2\right)$ is assumed the general free-phonon propagator, which is connected to Matsubara-domain phonon propagator as $D_{\mathbf{q}}^{(0)}\left(i \omega_n\right)=\int_0^\beta d \tau e^{+i \omega_n \tau} D_{\mathbf{q}}^{(0)}\left(\tau=\tau_1-\tau_2\right)=-\frac{2 \hslash \omega_{\mathbf{q}}}{\omega_n^2+\left(\hslash \omega_{\mathbf{q}}\right)^2} \quad$, $\omega_{\mathbf{q}}$ is the phonon dispersion or phonon frequency corresponding to the wavevector $\mathbf{q}$. ${f}_{\theta}$ here refers to the averaged embedded feature of $f_{l\theta}$ under the embedded form. In the actual physical expression of the neutron scattering, it should be reduced to the neutron scattering length. Here $M_{\alpha, Davg}$ and $\omega_{\alpha, Davg, \mathbf{q}}$ are the ensemble averaged mass and frequency for the $\alpha$ type defects after the impurity average.  The impurity averaged frequency may bring subtlety to the response function and the final phonon density of state. However, the overall contribution is explicitly related to the defect concentration and its associated defect feature embeddings. We emphasize that this expression is a formal approximation, not a physically exact description of defect vibrations. It serves to provide a structured mathematical form that supports reasoning and feature construction in our ML model, rather than modeling defect dynamics with full physical accuracy. The possible actual subtle non-linear contribution should be modelled through ML. 

Finally, notice that the imaginary part of $\chi _{\rho \rho }^R({\bf{q}},E) = \hslash {\chi _{\rho \rho }}({\bf{q}},i{\omega _n} \to E + i\delta )$ corresponds to the spectral function, whose momentum integration yields the phonon density of states (PDoS) $D(E)$:
\begin{equation}
    D(E)=\int A({\bf{q}}, E) d \mathbf{q} = -2\int \text{Im} \chi _{\rho \rho }^R({\bf{q}},E)d \mathbf{q}
    \label{PDoS-IAC}
\end{equation}
and for the sole defect contribution, we should have the embedding
\begin{equation}
\begin{split}
    D(E)-D_0(E) \xrightarrow{\text{Embedding}}   &-2\int \left(\text{Im} \chi _{PD}^R({\bf{q}},E)+\text{Im} \chi _{DD}^R({\bf{q}},E)\right)d \mathbf{q} \\&= \int \lambda(\mathbf{q})\sum_{\alpha,\beta} n_{\alpha} n_{\beta} g_{\alpha\theta} g_{\beta\theta}\text{d}\mathbf{q} + \text{Defect-pristine cross terms}
\end{split}
\label{eq:retarded}
\end{equation}
where $D(E)$ is the defective PDoS, $D_0(E)$ represents the pristine PDoS, $\lambda(\mathbf{q})$ is the prefactor with $\mathbf{q}$ dependence. $\chi _{PD}^R({\bf{q}},E)$ and $\chi _{DD}^R({\bf{q}},E)$ are the corresponding embeded retarded response function of $\chi _{PD}({\bf{q}},i{\omega _n})$ and $\chi _{DD}({\bf{q}},i{\omega _n})$, respectively. Thus, there exists a direct link from the defect embeddings -- which is defined in IAC -- and the measured PDoS $D(E)$. 
This connection confirms the intuition that PDoS is inherently sensitive to the type and distribution of dopants present in the material.

\section{Training Data and Architecture for DefectNet}

\subsection{Statistics of training data}
\begin{figure}[!htbp]
  \centering
  \includegraphics[width=1\textwidth]{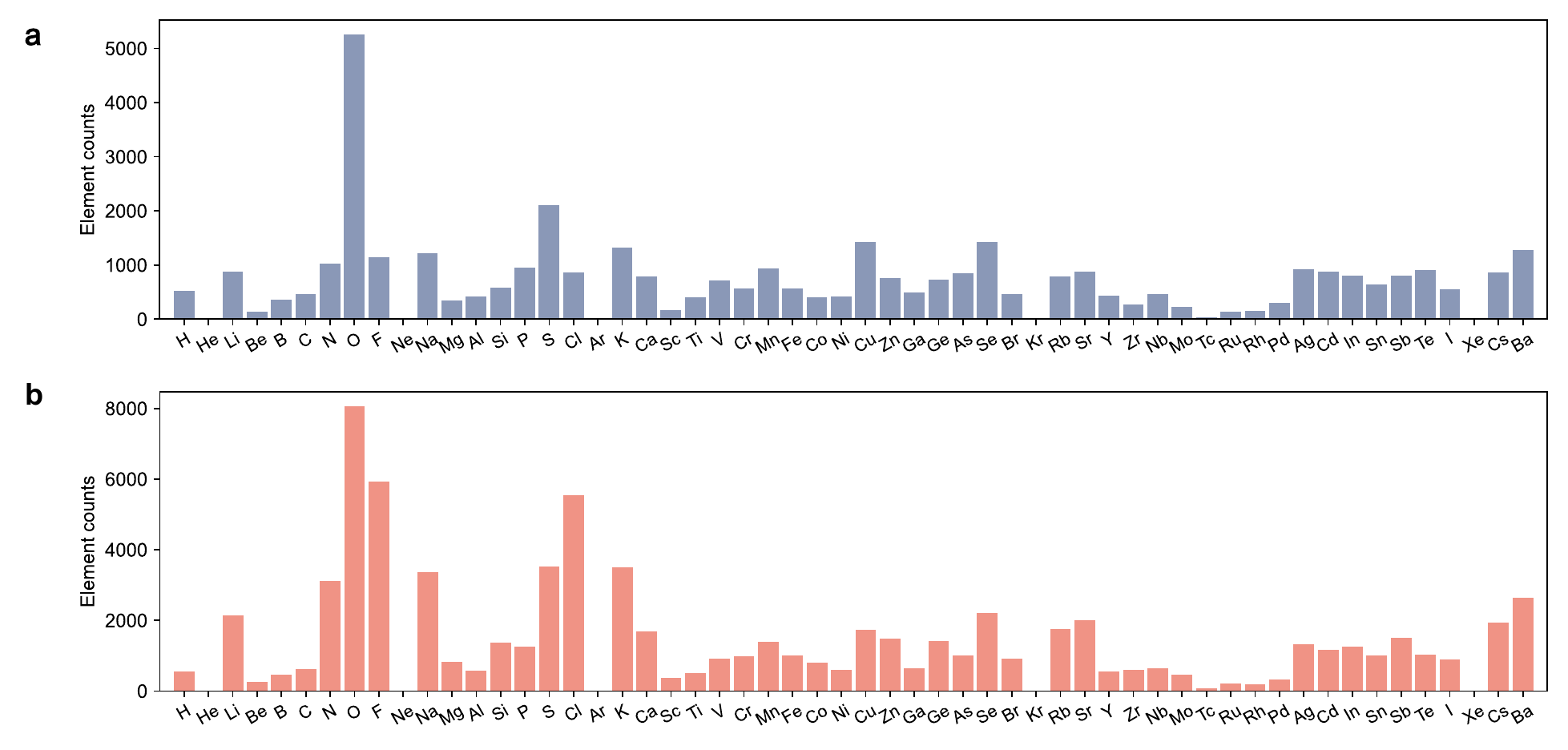}
  \caption{\textbf{Elemental distributions in the DefectNet training dataset.} \textbf{a.} Elemental composition of pristine (undoped) crystals. \textbf{b.} Elemental composition of doped crystals.}
  \label{fig-elements}
\end{figure}

In this section, we analyze the elemental composition and dopant relationships within the DefectNet training dataset. Elemental distributions of pristine (undoped) and doped crystals are illustrated in Fig. \ref{fig-elements}. In pristine crystals, oxygen (O), sulfur (S), and selenium (Se) emerge as the most abundant elements, whereas doped crystals predominantly contain oxygen (O), fluorine (F), and chlorine (Cl). Notably, among the 56 elements considered, all five noble gases (He, Ne, Ar, Kr, Xe) are absent from the dataset, consistent with their limited presence in stable crystalline materials.

\begin{figure}[!htbp]
  \centering
  \includegraphics[width=0.85\textwidth]{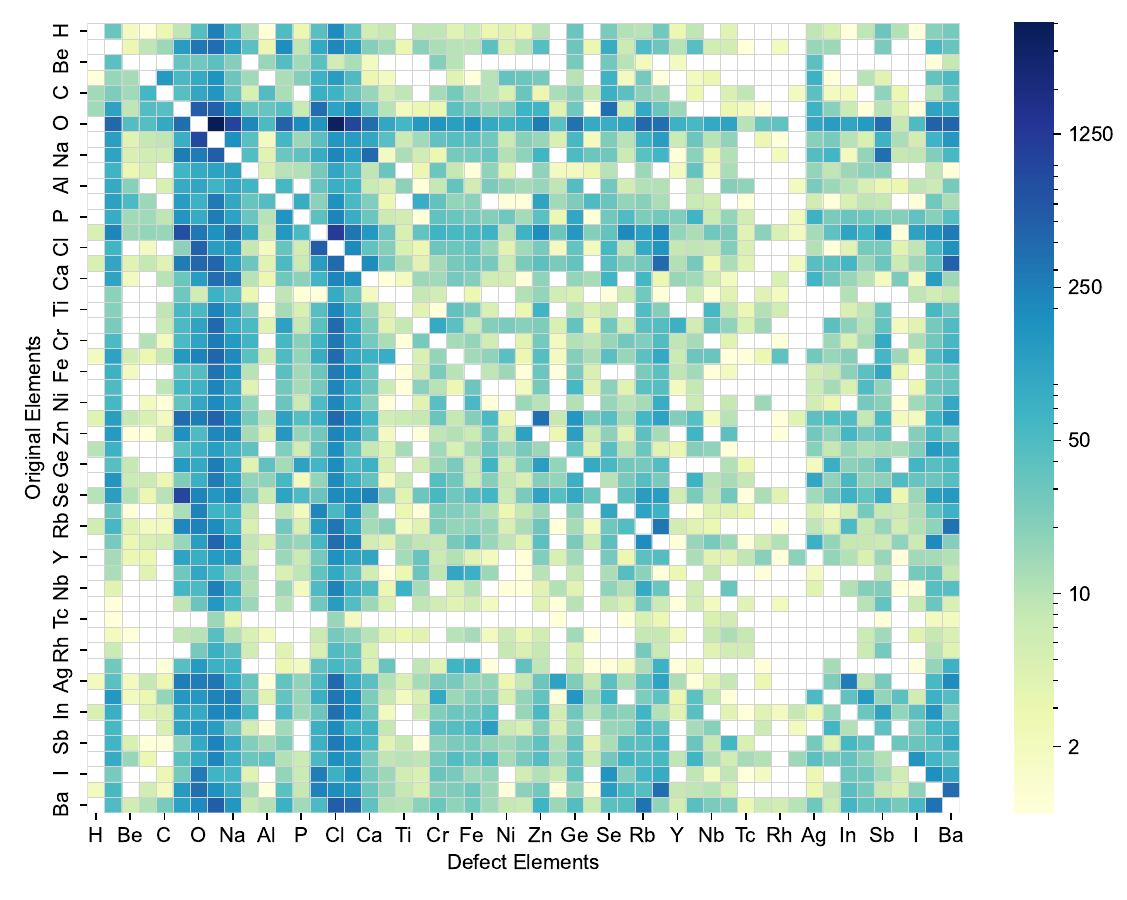}
  \caption{\textbf{Heatmap visualization of defect element frequencies in doped crystalline materials among the DefectNet dataset.} Rows represent 56 original host elements, and columns represent 56 defect-introduced elements (part of chemical elements are omitted in the label). }
  \label{fig-correlation}
\end{figure}

In Fig. \ref{fig-correlation}, we demonstrate a heatmap visualization of defect element frequencies in doped crystalline materials among the DefectNet dataset.  Each cell quantifies the frequency with which a given host element (rows) is replaced by a different dopant element (columns), visualized on a logarithmic scale to accommodate the wide dynamic range of substitution events. The diagonal entries representing self-substitution are naturally zero. Notably, several substitution trends emerge from the off-diagonal structure. Halogen elements such as F, Cl, and alkaline earth metals such as Na, K, and Ca exhibit broad substitution profiles, consistent with their high mobility and prevalence in defect chemistry. Sparse regions of the matrix highlight chemically unfavorable or structurally constrained substitutions, often involving large size or charge mismatches heavy post-transition metals. 

\subsection{Architecture of DefectNet}

\begin{figure}
  \centering
  \includegraphics[width=1.0\textwidth]{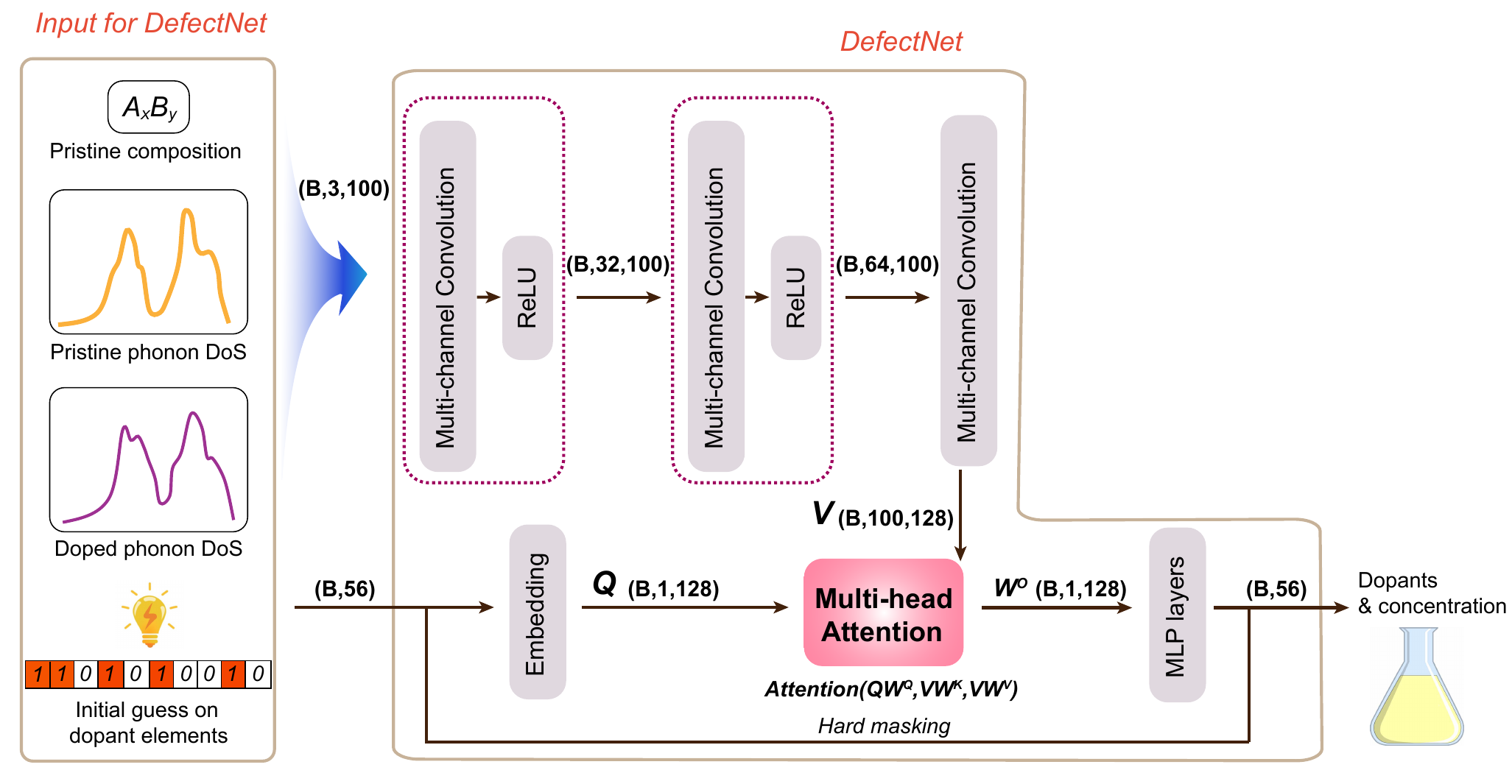}
  \caption{\textbf{Detailed architecture of the DefectNet model with annotated tensor dimensions.} The model takes as input three aligned one-dimensional signals of length 100: the pristine PDoS, the doped PDoS, and a vector encoding the parent crystal composition. These are concatenated into a 3-channel input tensor of shape ($B$,3,100), where $B$ is the batch size. The spectral encoder processes this input through multiple 1D convolutional layers and ReLU activations, yielding a latent feature sequence of shape ($B$,100,128), denoted as the value matrix $V$. Simultaneously, the initial dopant guess is represented as a binary vector of length 56, which is embedded in shape ($B$,1,128) to form the query matrix $Q$.  The attention output is then passed through a multi-layer perceptron to predict dopant concentrations, while a masking step is applied to ensure only the initially guessed dopants receive non-zero predictions. All tensor dimensions shown reflect batch-wise shapes.}
  \label{fig-Architecture}
\end{figure}

Here we illustrate the detailed neural network architecture of the proposed DefectNet. The DefectNet model is implemented in PyTorch and follows a modular architecture consisting of four components: (1) a spectral encoder based on 1D convolutions, (2) a dopant embedding module, (3) a multi-head attention mechanism, and (4) a dopant-masked module, which will be discussed the next subsection. This section details the first three components, as depicted in Fig.\,\ref{fig-Architecture}.

Each training instance consists of three aligned one-dimensional inputs of fixed length $L=100$ for the normalized PDoS, including the PDoS of the pristine host material and the PDoS after doping. Besides, we input a zero-padded vector encoding the host composition as well. These inputs are concatenated along a channel axis and processed through a sequence of one-dimensional convolutional layers with ReLU activations. The convolutional stack extracts local spectral features and transforms the input into a latent representation with $d=128$ channels, corresponding to an embedding dimension of 128. The result is a sequence of 100 embedded spectral tokens capturing both local and global vibrational characteristics.

In parallel, the initial guess of dopant candidates is provided as a 56-dimensional binary vector, indicating which dopants are considered for a given sample. This binary vector is projected into the same 128-dimensional latent space through a fully connected layer, resulting in a single dopant embedding vector. This embedding serves as a global query for the attention mechanism, allowing the model to modulate its spectral interpretation based on the specific dopant context.

To model the interaction between the dopant candidates and vibrational features, DefectNet employs a multi-head self-attention mechanism \cite{vaswani2017attention}. The dopant embedding acts as the query $Q$, while the spectral features $V$ act as both the key and value matrices. The attention follows the standard scaled dot-product formulation:
\begin{equation}
\mathrm{Attention}(Q, K, V) = \mathrm{softmax}\left( \frac{QK^\top}{\sqrt{d_k}} \right) V,
\end{equation}
where $Q \in R^{B\times 1 \times d}$, $K = V \in R^{B \times L \times d}$.
To capture diverse spectral-dopant relationships, DefectNet uses multi-head attention:
\begin{equation}
\mathrm{MultiHead}(Q, V, V) = \mathrm{Concat}(h_1, h_2, \ldots, h_H) W^O,
\end{equation}
where each attention head is computed as:
\begin{equation}
h_i = \mathrm{Attention}(QW_i^Q, VW_i^K, VW_i^V),
\end{equation}
and $W_i^Q$, $W_i^K$, $W_i^V$, and $W^O$ are learnable weight matrices specific to each head.
This attention mechanism enables the model to identify which parts of the vibrational spectrum are most informative for determining the presence and identity of dopants. By aligning spectral patterns with dopant-specific context, the model can resolve subtle or overlapping features arising from co-doping or low-concentration defects.

The output of the attention block is a 128-dimensional feature vector that summarizes dopant-aware spectral information. This vector is passed through a multi-layer perceptron consisting of two hidden layers and a final regression layer that outputs predicted concentration values for all 56 possible dopants. 

\subsection{Dopant recommender and hard masking}

To ensure both user-friendly deployment of DefectNet and physical consistency with dopant selection, the model’s output is element-wise masked using the original dopant vector. There are two typical use cases for dopant selection. In the first case, users have prior knowledge of the dopant species of interest. They can apply a hard masking operation to enforce predictions only on the specified elements. In the second case, users may be uncertain about the exact dopants present in the material system. In this scenario, a dopant recommender can be used to construct an appropriate hard masking, thereby restricting the predictions to a physically plausible set of dopant elements.

\begin{figure}[!htbp]
  \centering
  \includegraphics[width=0.8\textwidth]{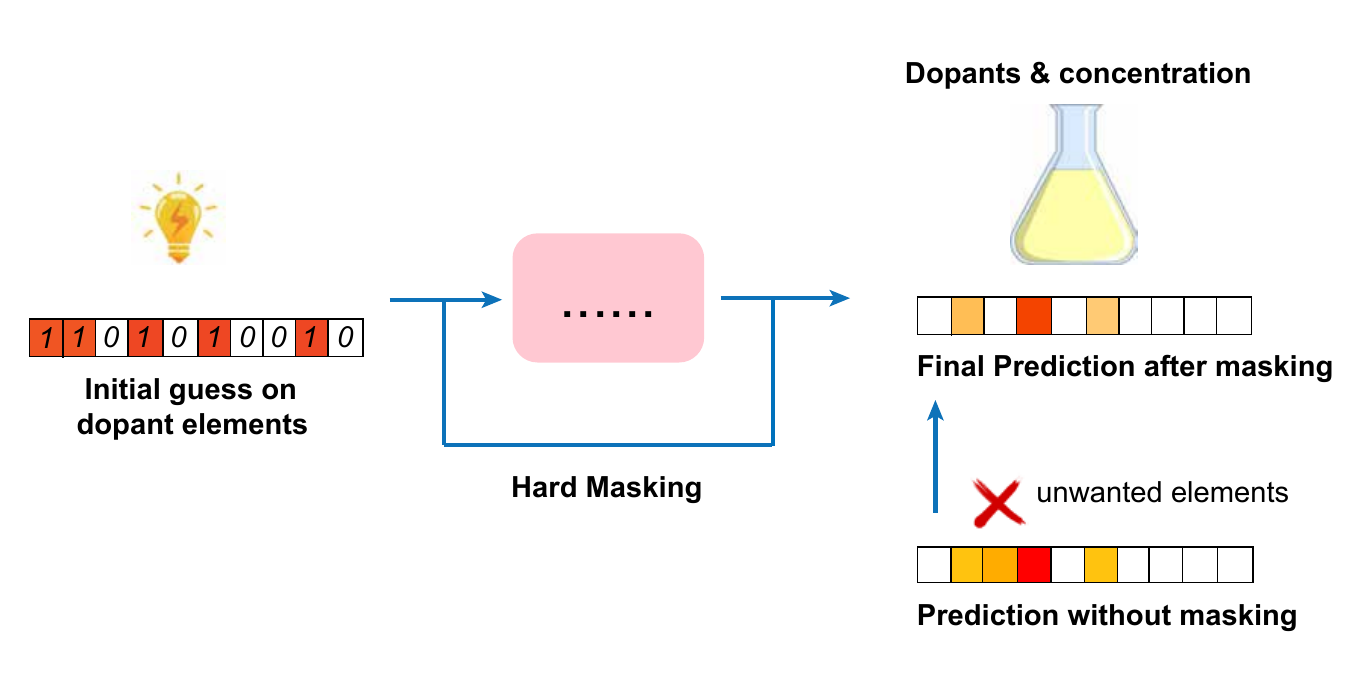}
  \caption{\textbf{Hard masking procedure in DefectNet.} This masking is employed to enforce that only selected dopants receive non-zero predictions. Such a design allows flexibility and alignment with the user's initial guess on dopant species.}
  \label{fig-Masking}
\end{figure}

The dopant recommender, which evaluates the compatibility between candidate dopant elements and the host material, has been described in the main text. The hard masking mechanism, as illustrated in Fig.\,\ref{fig-Masking}, imposes an explicit constraint whereby only the initially selected dopants are allowed to receive non-zero predictions. Specifically, the mean squared error loss is computed only over those dopant elements selected in the initial guess. This loss masking reinforces the effect of the architectural masking by ensuring that the model is only penalized for predicting concentrations on plausible dopant species, which further improves training stability, prevents overfitting to distractor classes, and aligns optimization with the physical priors embedded in the dopant recommender.

\section{Additional Results for DefectNet Prediction}
In this section, we present additional prediction results of the pre-trained DefectNet model.

\subsection{Prediction within the full dataset}

\begin{figure}[!htbp]
  \centering
  \includegraphics[width=1\textwidth]{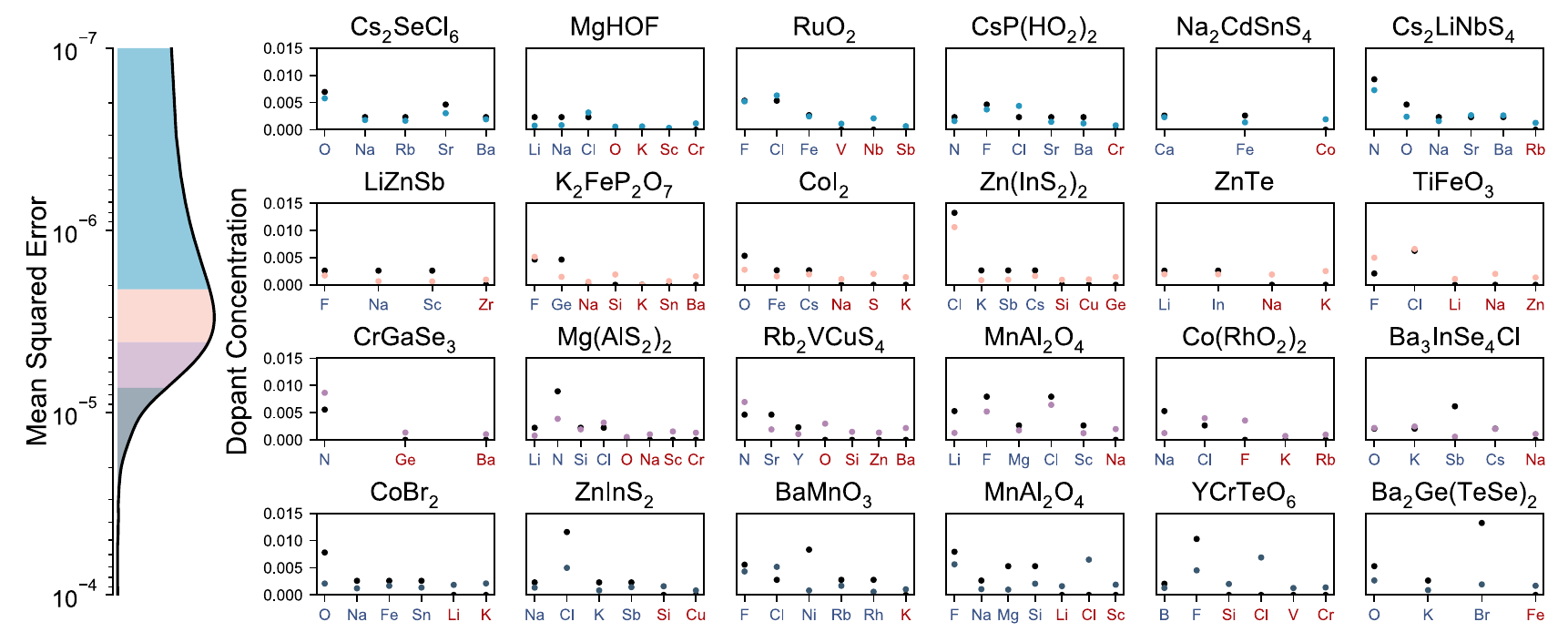}
  \caption{\textbf{Additional prediction results of DefectNet on defects, with all PDoS curves randomly mixed and split to the testing set.} 
  Each panel compares the predicted (colored) and ground-truth (black) dopant concentrations across the full dataset with multiple coexisting dopants.
  Additionally, distractor dopant candidates (elements colored red) exist during the testing process.
  The leftmost plots display the distribution of mean squared error (MSE) in predicted defect concentrations, grouped into quartiles based on overall model performance sorted by error (top to bottom: low to high).}
  \label{fig-Full_dataset_1}
\end{figure}

\begin{figure}[!htbp]
  \centering
  \includegraphics[width=1\textwidth]{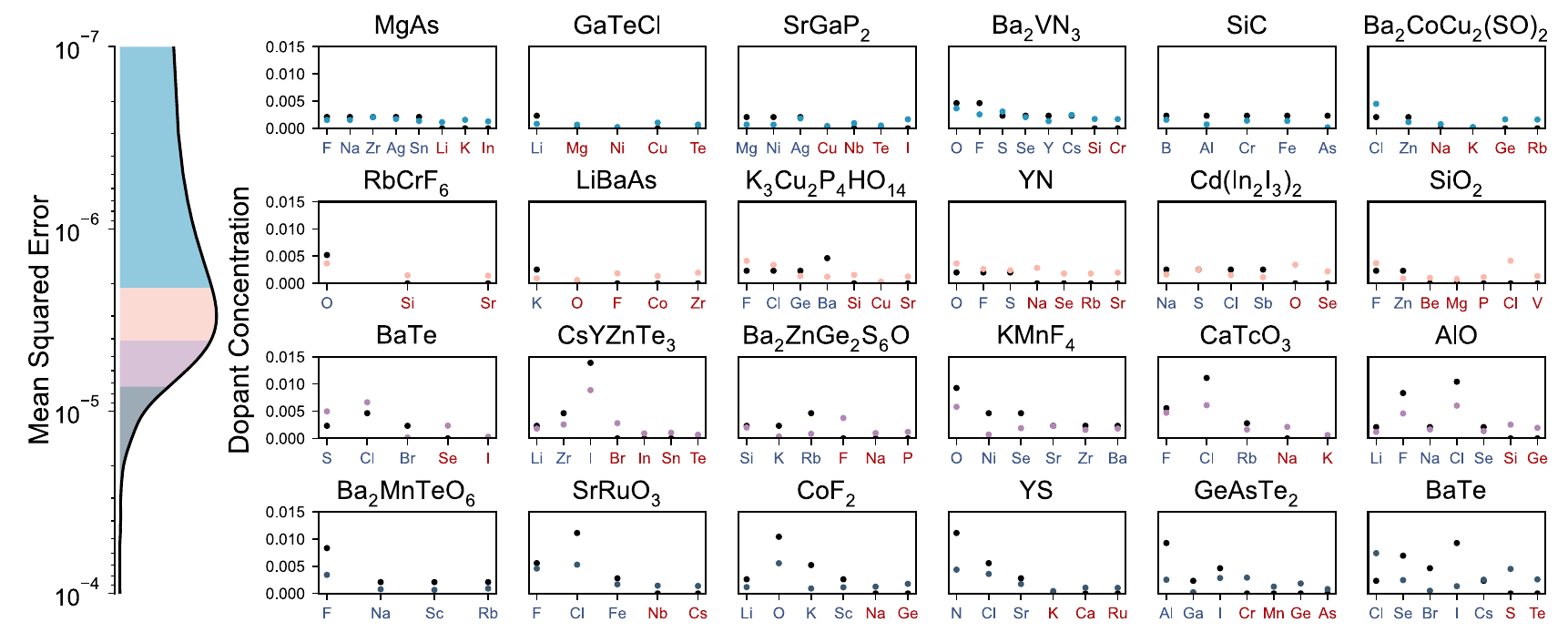}
  \caption{\textbf{Additional prediction results of DefectNet on defects, with all PDoS curves randomly mixed and split to the testing set.} Continued from Fig.\,\ref{fig-Full_dataset_1}.}
  \label{fig-Full_dataset_2}
\end{figure}

In Fig.\,\ref{fig-Full_dataset_1} and Fig.\,\ref{fig-Full_dataset_2}, the full dataset comprising parent crystals and their corresponding doped variants is randomly partitioned into distinct training and testing subsets. This setup constitutes an in-distribution scenario typical of standard machine learning validation protocols, where spectral features and dopant interactions present in the test set closely resemble those encountered during training. Within this evaluation framework, DefectNet demonstrates robust and highly accurate predictive capabilities, effectively capturing complex spectral signatures across diverse dopant configurations. 

\newpage
\subsection{Prediction extrapolating parent crystals beyond the dataset}

\begin{figure}[!htbp]
  \centering
  \includegraphics[width=1\textwidth]{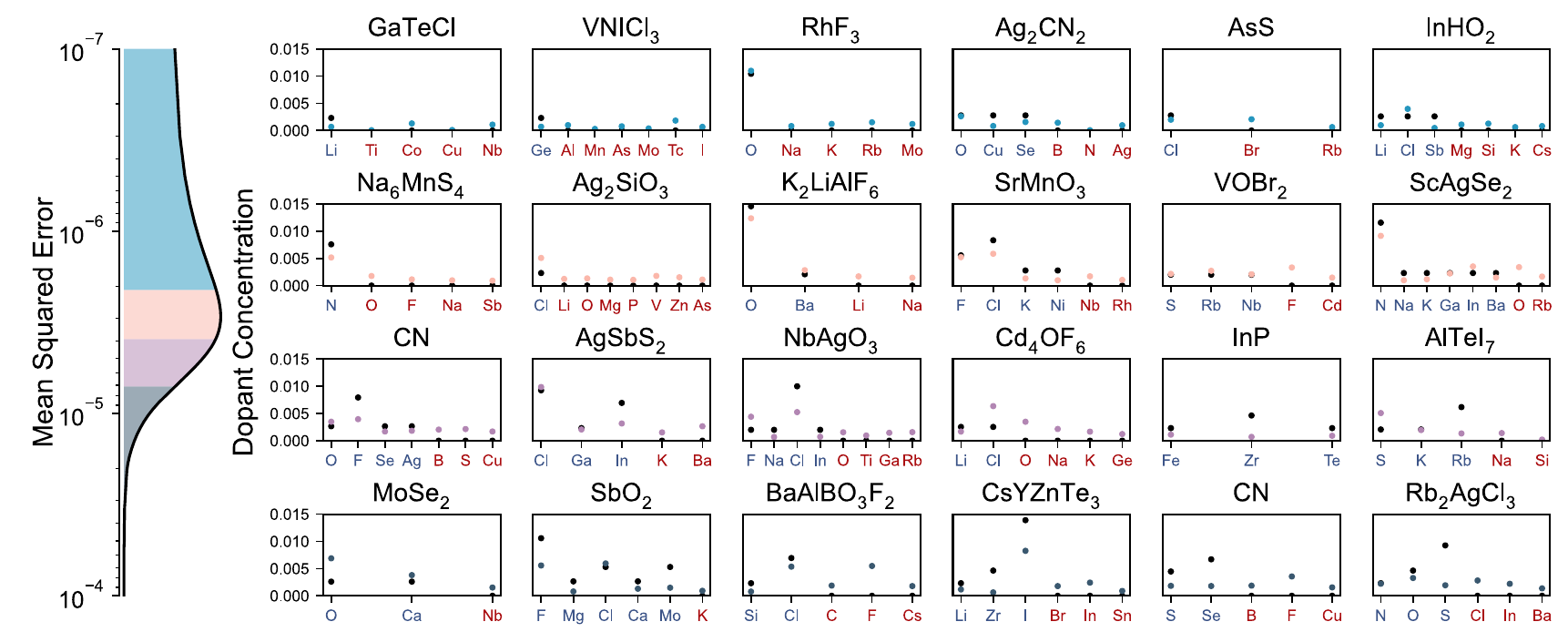}
  \caption{\textbf{Additional prediction results of DefectNet on defects, with a fraction of parent crystals held out entirely from training to test DefectNet's ability to extrapolate to unseen materials.} Each panel compares the predicted (colored) and ground-truth (black) dopant concentrations across the full dataset with multiple coexisting dopants.
  Additionally, distractor dopant candidates (elements colored red) exist during the testing process.
  The leftmost plots display the distribution of mean squared error (MSE) in predicted defect concentrations, grouped into quartiles based on overall model performance sorted by error (top to bottom: low to high).}
  \label{fig-Extrapolation_1}
\end{figure}

\begin{figure}[!htbp]
  \centering
  \includegraphics[width=1\textwidth]{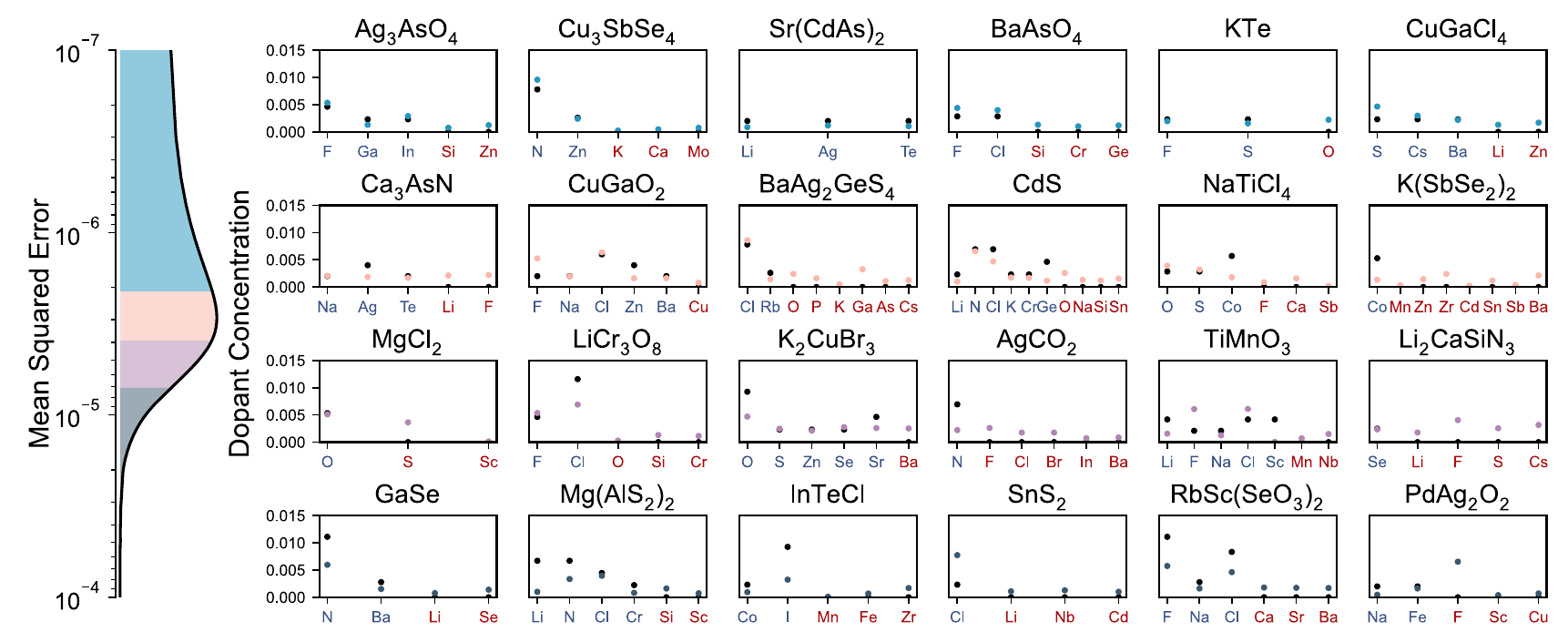}
  \caption{\textbf{Additional prediction results of DefectNet on defects, with a fraction of parent crystals held out entirely from training to test DefectNet's ability to extrapolate to unseen materials.} Continued from Fig.\,\ref{fig-Extrapolation_1}. }
  \label{fig-Extrapolation_2}
\end{figure}

DefectNet achieves state-of-the-art predictive performance even when evaluated under extrapolative conditions that rigorously test its generalization capabilities. In Fig.\,\ref{fig-Extrapolation_1} and Fig.\,\ref{fig-Extrapolation_2}, a subset of parent crystals is completely excluded from the training process, ensuring that no such undoped structures are present during model training. Despite the absence of prior exposure to these parent compounds, DefectNet continues to deliver highly accurate predictions.

\newpage
\section{Validation on Experimental Data}
\subsection{Experimental fine-tuning on SiGe alloy}
\begin{figure}[!htbp]
  \centering
  \includegraphics[width=0.7\textwidth]{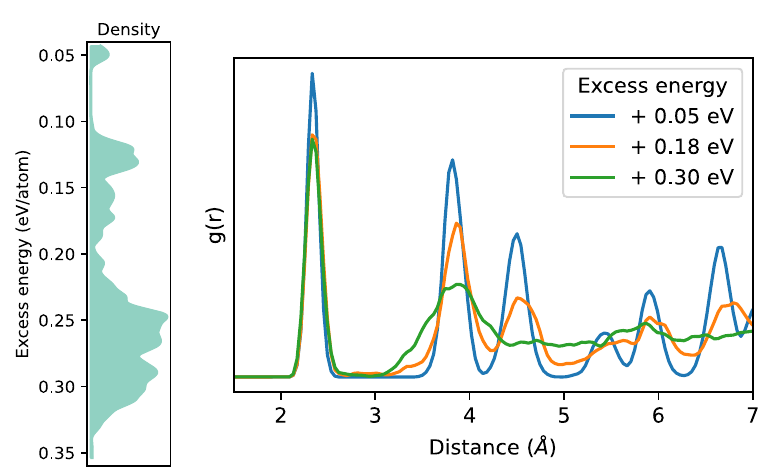}
  \caption{\textbf{Statistics for the training data used for fine-tuning DefectNet.} Left panel: distribution of excess energies of the sampled bulk Si structures. Right panel: radial distribution functions $g(r)$ for three representative energy levels, indicating increasing structural disorder at higher excess energies.}
  \label{fig-aSigr}
\end{figure}

As discussed in the main text, we have demonstrated the feasibility of applying DefectNet to SiGe alloys.
Here we show additional structural statistics of the training dataset, which consists of a diverse set of disordered bulk silicon configurations used for fine-tuning DefectNet.
As shown in Fig.\,\ref{fig-aSigr}, structures with higher excess energy exhibit increasingly disordered local environments, as evidenced by the progressive loss of long-range order in their radial distribution functions $g(r)$.
The dataset spans a broad range of structural motifs, from nearly crystalline to amorphous-like, enabling comprehensive sampling of the disordered configuration space. Ge atoms are subsequently doped into these host structures to construct the full training set for DefectNet.

\subsection{Experimental fine-tuning on Al-doped MgB$_2$}
\begin{figure}[!htbp]
  \centering
  \includegraphics[width=0.6\textwidth]{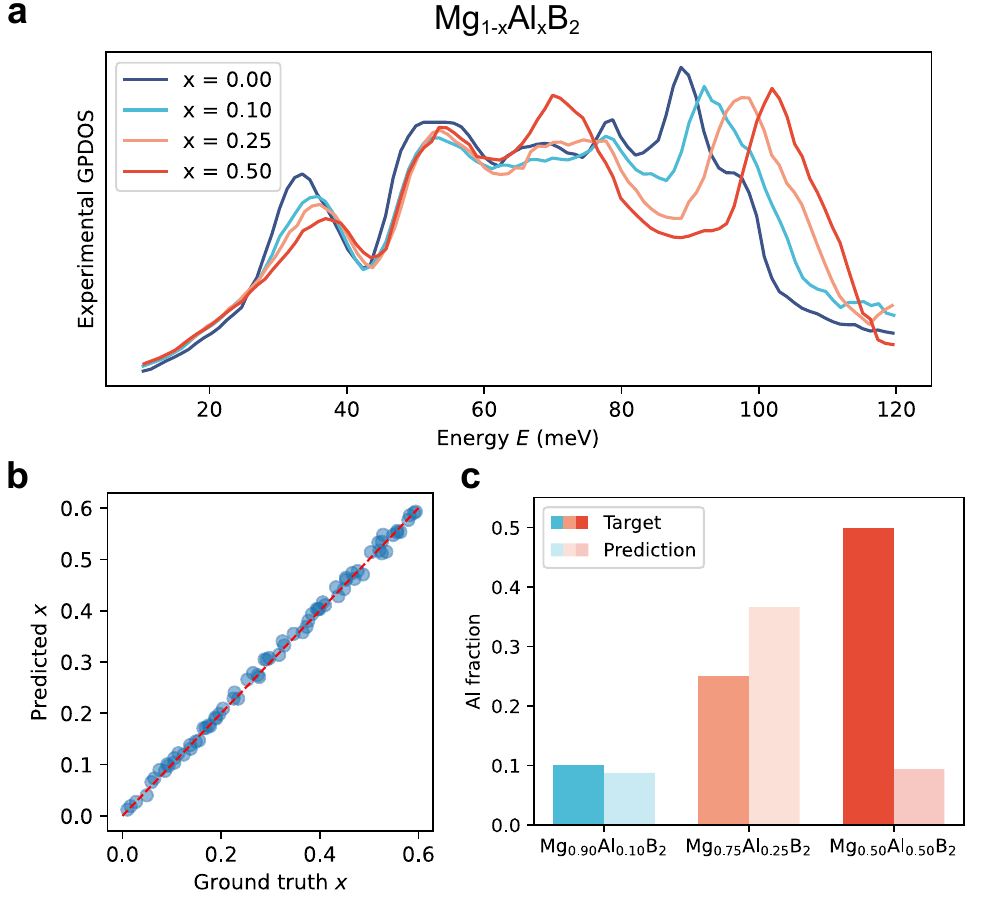}
  \caption{\textbf{Fine-tuning DefectNet and validation on experimental Al-doped MgB$_2$.} \textbf{a.} Experimental generalized phonon density of states (GPDoS) curves for bulk MgB$_2$ and Mg$_{1-x}$Al$_x$B$_2$ with $x=0.10$, $0.25$, and $0.50$, as is extracted from Ref.\,\cite{yokoo2004evidence}. 
  \textbf{b.} Scattered plot for predicted versus ground-truth Al concentration $x$. \textbf{c.} Comparison between DefectNet-predicted Al fractions $x$ and the lab reported ground-truth for the three Mg$_{1-x}$Al$_x$B$_2$ samples.
  }
  \label{fig-mgb2}
\end{figure}

To evaluate the transferability of our defect learning framework beyond simulated datasets, in addition to the SiGe alloy presented in the main text, we also fine-tune DefectNet on experimental data for the Al-doped superconductor Mg$_{1-x}$Al$_x$B$_2$, as is shown in Fig.\,\ref{fig-mgb2}. 
This system is a well-known electron-doped superconductor where Al substitution modifies both electronic and vibrational properties, providing a stringent test for spectral-based ML prediction for superconducting systems other than semiconductors.
We collect experimental generalized phonon density of states (GPDoS) for various Al concentrations ($x = 0$, $0.10$, $0.25$, $0.50$) as reported in Ref.\cite{yokoo2004evidence} and reproduced in Fig.\,\ref{fig-mgb2}(a). The DefectNet model is fine-tuned on simulated spectra of pristine and Al-doped MgB$_2$ using the MatterSim v1.0.0-1M MLIP.
After fine-tuning, the model achieved a root-mean-square error (RMSE) of 0.008 on synthetic data (Fig.\,\ref{fig-mgb2}(b)). 

Predictions for dopant concentration $x < 0.25$ showed reliable agreement with experimental labels, as shown in Fig.\,\ref{fig-mgb2}(c). However, at a higher doping level ($x=0.50$), prediction performance degraded significantly.
This breakdown likely originates from the increasing complexity and nonlinearity of the experimental signal at high dopant levels, which diverges from the cleaner and more idealized training spectra generated from simulations. 
Moreover, the discrepancy between clean DFT-generated spectra, MLIP-based approximations, and actual measured spectra introduces a domain shift that is not fully captured by the current model. These findings underscore both the promise and current limitations of purely spectrum-driven approaches for defect inference in complex materials and highlight opportunities for incorporating domain adaptation techniques or hybrid physical priors to improve robustness.

\newpage
\bibliography{refs.bib}